\documentclass[11pt]{article}
\usepackage{epsf}
\usepackage{graphicx}
\usepackage{amssymb}
\usepackage{amsmath}

\long\def\comment#1{ }    

\newcommand{\slL}{\raise.15ex\hbox{$/$}\kern-.53em\hbox{$L$}}
\newcommand{\slP}{\raise.15ex\hbox{$/$}\kern-.67em\hbox{$P$}}
\newcommand{\slp}{\raise.1ex\hbox{$/$}\kern-.63em\hbox{$p$}}
\newcommand{\slq}{\raise.1ex\hbox{$/$}\kern-.63em\hbox{$q$}}
\newcommand{\slv}{\raise.1ex\hbox{$/$}\kern-.63em\hbox{$v$}}
\newcommand{\slR}{\raise.15ex\hbox{$/$}\kern-.53em\hbox{$R$}}
\newcommand{\slQ}{\raise.15ex\hbox{$/$}\kern-.53em\hbox{$Q$}}
\newcommand{\slK}{\raise.15ex\hbox{$/$}\kern-.53em\hbox{$K$}}
\newcommand{\slk}{\raise.15ex\hbox{$/$}\kern-.53em\hbox{$k$}}
\newcommand{\slSigma}{\raise.15ex\hbox{$/$}\kern-.53em\hbox{$\Sigma$}}
\newcommand{\slcalP}{\raise.15ex\hbox{$/$}\kern-.63em\hbox{$\cal P$}}
\newcommand{\slA}{\raise.15ex\hbox{$/$}\kern-.73em\hbox{$A$}}
\newcommand{\slbfA}{\raise.15ex\hbox{$/$}\kern-.73em\hbox{${\imb A}$}}
\newcommand{\slpartial}{\raise.15ex\hbox{$/$}\kern-.53em\hbox{$\partial$}}

 \def\pt{$p_{_T}$}
 
  \def\pt3{{p_{_T{_3}}}}
\def\pt4{{p_{_T{_4}}}}

\def\gsim{\mbox{~{\raisebox{0.4ex}{$>$}}\hspace{-1.1em}
{\raisebox{-0.6ex}{$\sim$}}~}}
\def\lsim{\mbox{~{\raisebox{0.4ex}{$<$}}\hspace{-1.1em}
{\raisebox{-0.6ex}{$\sim$}}~}}

\newcommand{\be}{\begin{equation}} \newcommand{\ee}{\end{equation}}
\newcommand{\bea}{\begin{eqnarray}} \newcommand{\ena}{\end{eqnarray}}
\long\def\comment#1{ }    

\def\lsim{\raise0.3ex\hbox{$<$\kern-0.75em\raise-1.1ex\hbox{$\sim$}}}
\def\gsim{\raise0.3ex\hbox{$>$\kern-0.75em\raise-1.1ex\hbox{$\sim$}}}

\begin{document}

\begin{titlepage}
\begin{flushright}
           \today\\ hep-ph/xxyynnn\\ LPTHE-05-14\\ LAPTH-1092/05 \\
	   IMSc-2005/02/03\\IISc-CHEP/02/05 
\end{flushright}

\vskip 1.5cm {\large \centerline{\bf Deep inelastic scattering and 
forward $\pi^0$ production at NLO}}
\vskip 1cm \centerline{ P. Aurenche$^{a}$,
Rahul Basu$^{b}$, M. Fontannaz$^{c}$, R.M.~Godbole$^{d}$}

\vskip .5cm

{\small\sl \centerline{ $^a$Laboratoire d'Annecy-le-Vieux de Physique Th\'eorique
LAPTH{{\footnote{UMR 5108 du CNRS, associ\'ee \`a l'Universit\'e de
Savoie}}},}

\centerline{B.P.110, F-74941 Annecy-le-Vieux Cedex, France}

\vskip .5cm

\centerline{ $^b$The Institute of Mathematical Sciences,}

\centerline{Chennai 600 113, India}

\vskip .5cm

{\centerline{ $^c$Laboratoire de Physique Th\'eorique, UMR 8627 CNRS,}

\centerline{Universit\'e Paris XI, B\^atiment 210, F-91405 Orsay Cedex, France}

\vskip .5cm

\centerline{ $^d$Center for High Energy Physics,}

\centerline{Indian Institute of Science, Bangalore 560 012, India}}}

\vskip 1cm
\begin{abstract}  
We present a detailed phenomenological study of forward hadron ($\pi^0$)
production in deep inelastic scattering, with both the direct and the resolved 
contributions calculated to NLO accuracy. A comparison of the 
theoretical predictions for the various distributions with the H1 data and 
a  study of stability of the QCD predictions under changes of scales is  
the focus of this study. We obtain a very good overall description of the 
recent H1 data with the choice of scale $Q^2 + E^2_\bot$, in contrast to the
$(Q^2 + E^2_\bot)/2$ required earlier when the resolved contribution
was included only at LO accuracy. We find a more modest variation of the 
predictions, as the scale is changed from $(Q^2 + E^2_\bot)/2$ to 
$2(Q^2 + E^2_\bot)$, as compared to the case where the resolved contribution
was included only at LO accuracy. This variation is  of the order of the 
rather large experimental errors. Unfortunately, this fact prevents us from 
concluding that perturbation theory gives an unambiguous prediction for 
forward particle production in deep inelastic scattering. However, the overall
success of perturbative QCD in explaining the small $x_{Bj}$  data means that
perhaps a full resummation of the BFKL ladder is not called for. We notice the 
need for rather large resolved contributions to explain the data at 
low $x_{Bj}$ even at somewhat larger $Q^2$ values.
\end{abstract}

\end{titlepage}

\newpage

\section{Introduction}

Recent experimental data, from the H1 collaboration~\cite{h1-2004}, on large
transverse energy hadron production in deep-inelastic scattering have generated
several theoretical papers attempting to explain the data within the framework
of perturbative Quantum Chromodynamics in the next-to-leading order (NLO)
approximation. These experimental results confirm and extend older data from
H1~\cite{h1-1999,h1<1999} and ZEUS~\cite{zeus-1995,zeus-1999}. Since these H1 
and ZEUS data
on forward hadrons as well as the data on forward 
jet~\cite{Aktas:2004px,Breitweg:1998ed,Breitweg:2000sv} 
production at large transverse momentum probe the small Bjorken-$x_{Bj}$
region, it was argued \cite{4r} that they would be  ideally suited to probe the
Balitsky-Fadin-Kuraev-Lipatov~\cite{bfkl} (BFKL) regime, where the resummation
of $\ln(1/x_{Bj})$ terms is important, and that they would show the breakdown
of the Dokshitzer-Gribov-Lipatov-Altarelli-Parisi~\cite{dglap} (DGLAP) regime. 
In this respect the single hadron data are more relevant than the jet data
since they cover a lower Bjorken-$x_{Bj}$ range, down to $x_{Bj} = 4.
10^{-5}$~\cite{h1-2004}. These data are also expected to be more accurate than
the jet data because of the difficulty of jet identification at low transverse
momentum and in the forward region.

A comparison of older H1~\cite{h1-1999} results on single $\pi^0$ production,
with a model based on lowest order matrix elements and parton 
cascades~\cite{lepto}, shows 
very strong disagreement between data and theory. The model falls much below 
the data (a factor 5 to 10 at low $Q^2$ and low $x_{Bj}$) and, besides, the
shape of the $x_{Bj}$  dependence is incorrect. Adding the contribution where
the virtual photon is resolved~\cite{rapgap} reduces somewhat the disagreement
at large $Q^2$ but falls short of the data at small $Q^2$ unless a very large
scale is chosen in the evaluation of the anomalous photon
component~\cite{h1-2004}. However, predictions based on improved
leading order BFKL dynamics~\cite{kwiecin} show a better overall agreement when
compared to H1 data, specially at low $Q^2$, but they do not describe the $Q^2$
evolution correctly~\cite{h1-2004}.

The recent theoretical developments concern mainly the calculation of the
single hadron production in the NLO
approximation~\cite{Daleo:2003xg,Daleo:2004pn,Aurenche:2003by,Kniehl:2004hf,Fontannaz:2004ev}.
Higher order diagrams neglected in the earlier approaches modify the picture in
several ways. They generate new topologies and new hard scattering processes
which should be considered as new  Born terms. For example, at the lowest order
(LO) the hard scattering terms are mediated by quark exchange while in the NLO
approximation  processes with gluon exchange appear and these become specially 
important~\cite{Aurenche:2003by} in the forward region where the presence of
the gluon pole enhances such terms. Also, at NLO, terms associated with the
$q{\bar q}$ collinear component of the virtual photon, which build up the
photon structure function (the so-called resolved component), appear. As is
well known, a ``large" logarithm arises, asymptotically of type $\ln(E_\bot^2 /
Q^2)$ when $E_\bot^2\gg Q^2$ ($E_\bot$ is the hadron transverse momentum in the 
$\gamma^*$-proton center of mass frame\footnote{The variable which we denote
$E_\bot$ here is called $p^*_T$ in Ref.~\cite{h1-2004}.}), and  
is associated with this structure function, specially when the photon virtuality
is small. This term can then be considered a leading order term although it
technically appears when calculating higher order diagrams
\cite{Drees:1995wh} . This is the reason
why it was introduced in~\cite{rapgap} where it indeed helped improve agreement
with the data. However, using a large scale in the photon structure function to
enhance the resolved contribution appears artificial. Indeed in an NLO
calculation, the increase of the Born resolved contribution is compensated by a
decrease of the higher order direct contribution, not included in
\cite{rapgap}, in such a way that the sum is more stable under changes of
scales.

It should be stressed that single hadron production in deep-inelastic
scattering experiments present a very stringent consistency test of
perturbative QCD and its various input distributions: it involves the proton
structure function as well as the hadronic fragmentation functions, all
quantities rather precisely measured in other experiments. As just discussed it
is also very sensitive to the virtual photon structure
function~\cite{Gluck:1994tv,Gluck:1999ub,Chyla:1999pw}, which is less well
known but which has been recently discussed in detail in~\cite{Fontannaz:2004ev}.

The common features of recent NLO
calculations~\cite{Daleo:2003xg,Daleo:2004pn,Aurenche:2003by,Kniehl:2004hf} are
the following. The cross section contains two (complicated) pieces: the
``direct" cross section, where the virtual photon couples directly to the hard
process,  and the ``resolved" cross section, where the photon acts as a
composite object which is a source of collinear partons taking part in the hard
subprocess. The direct contribution is calculated in the NLO approximation,
{\it i.e} up to ${\cal O}(\alpha^2_s$), while the cross section involving the
resolved component is calculated to lowest order (LO) accuracy with the photon scale
compensating term included in the higher order part of the direct piece. No
DGLAP type resummation is performed on the virtual photon structure function.
Using modern proton structure functions~\cite{mrst,cteq} and fragmentation
functions~\cite{kkp} a very good agreement is achieved with the data when using
a common (renormalization, factorization and fragmentation) scale set equal to
$(Q^2+E^2_\bot)/2$. However all the above calculations exhibit the same large 
scale dependence of the predictions mainly associated with the renormalization 
scale as will be
seen below. A rather large sensitivity of the predictions to the fragmentation
functions is also observed, with the  data clearly
favoring~\cite{Daleo:2004pn}, like other hadronic data, the parametrization of Kniehl, Kramer and
P\"otter~\cite{kkp} (KKP) over that of Kretzer~\cite{kretzer}.  Furthermore, in
\cite{Aurenche:2003by} a discussion is given to isolate the origin of the large
corrections terms and it is found that they are associated with processes with a
gluon exchange which are interpreted as the Born terms of the BFKL ladder. The
theoretical papers differ in the procedure to obtain the cross section: in
\cite{Daleo:2003xg,Daleo:2004pn} a calculation of the single particle spectrum
is performed with the infrared divergences compensated analytically while in
\cite{Aurenche:2003by,Kniehl:2004hf} a Monte-Carlo generator at the partonic
level is constructed with a numerical compensation of divergences.

In \cite{Fontannaz:2004ev}, the first evaluation, at the NLO accuracy, of the
resolved contribution is presented: it includes both the construction and the
use of the NLO virtual photon structure function as well as the NLO
calculation of the hard matrix elements for the resolved processes. In the
limited phenomenological analysis performed, good agreement with the data is
obtained with the scale $(Q^2+E^2_\bot)$, larger than that of the previous NLO
calculations. The importance of the resolved contribution to the cross section
is again emphasized and it is shown that its factorization scale dependence is 
reduced at the NLO accuracy compared to the LO calculation. It is then
expected that the full cross section will be less scale sensitive than in the
work of~\cite{Daleo:2003xg,Daleo:2004pn,Aurenche:2003by,Kniehl:2004hf}.

In the following we present a detailed phenomenological study of hadron
production in deep-inelastic scattering with both the direct and resolved contributions
calculated at NLO accuracy. A special emphasis will be put on the study
of the scale
variation of the cross section to determine the domain where the perturbative
QCD approach is reliable, {\it i.e.} stable under changes of scales.

In the next section we set up the theoretical framework and discuss the
instabilities related to the various scales introduced in the calculation
(factorization scale $M$ on the proton side and $M_\gamma$ on the photon side,
fragmentation scale $M_F$ and factorization scale $\mu$). A detailed comparison
with the various experimental H1 distributions~\cite{h1-2004} is performed
next: at small Bjorken-$x_{Bj}$  the large corrections are found to be related
to BFKL-like terms which appear in the NLO calculation in some approximation
(first corrections in $\alpha_s \ln(1/x_{Bj})$). Studying the $Q^2$ dependence
of the cross section will probe the photon structure function as it is expected
to play a dominant role at low $Q^2$ while at large $Q^2$ the direct term is
expected to dominate. Finally, studying the rapidity $\eta_\pi$ or
$x_\pi=E^{lab}_\pi/E^{lab}_p$ distributions, as well as the transverse momentum
distribution of the pion will help constrain the quark and gluon fragmentation
into pions.

Hadron production in DIS experiments offers a very rich structure: it is a two
scale problem, $Q^2$ and $E_\bot^2$, with a large variation in the ratio of
these two scales allowing the testing of the theoretical results in different
regimes. Furthermore, the small $Q^2$ limit makes it possible to
make contact with photoproduction experiments. Combining all the data will help
in understanding the non-perturbative input to the photon structure function
and its
decreasing importance when the virtuality of the photon increases. It will
give some insight on the transition from the non-perturbative to the 
perturbative regime, for the photon structure function.

\section{Theoretical framework}
\label{sec:theory}

We first discuss the features of the resolved cross section which were not
taken into account in the previous
papers~\cite{Daleo:2003xg,Daleo:2004pn,Aurenche:2003by,Kniehl:2004hf}. When
calculating the  higher order (HO) corrections to the direct contribution there
appear configurations where the virtual photon turns into an almost collinear
$q$-$\overline{q}$ pair with the quark or the antiquark subsequently
interacting with a parton from the proton. This HO contribution, in principle
negligible for $E_\bot^2$ close to $Q^2$, is important when $E_\bot^2$ becomes
large. In this  case we cannot content ourselves with the lowest order
expression of the quark distribution in the virtual photon, proportional to
$\ln (E_\bot^2 / Q^2)$. The latter must be replaced by a resummed LO or NLO
expression. The standard procedure consists in subtracting from the HO
corrections a term proportional to $\ln (M_{\gamma}^2 /Q^2)$ and to calculate a
resolved contribution with fully evolved parton distributions at the
factorization scale
$M_{\gamma}^2$. The $M_{\gamma}^2$-dependence of the resolved part is partly
compensated by the $\ln (E_\bot^2 / M^2_\gamma)$ counterterm which remains in
the direct HO contribution. In Ref. \cite{Fontannaz:2004ev} it was argued that 
a physical choice for the factorization scale is $M_{\gamma}^2 = (Q^2 + 
C_{\gamma}^2\  E_\bot^2)$ with $C_{\gamma}$ of order ${\cal O}(1)$.
With this scale, the resolved component is negligible when
$E_\bot^2 \ll Q^2$, whereas it is large when $E_\bot^2 \gg Q^2$. In
the latter case, we recover the standard factorization scale
$C_{\gamma}^2\  E_{\bot}^2$ of large-$E_\bot$ reactions.

Direct and resolved cross sections, calculated in the NLO approximation, have
been discussed respectively in Ref. \cite{Aurenche:2003by} and
\cite{Fontannaz:2004ev}. Here, we do not give the technical
details of these calculations which can be found in the relevant references,
but we summarize the main results that were obtained:

1) The HO corrections to the direct cross section are very large (in
the H1 kinematical domain) and essentially come from graphs containing
the exchange of a gluon in the $t$-channel. These graphs represent a
zeroth order approximation to the BFKL ladder. 

2) The NLO direct cross section strongly depends on the renormalization
scale $\mu$. 

3) With the factorization scale $M_\gamma^2= Q^2+E_\bot^2$ the NLO resolved 
contribution is as large as the NLO direct one.
Therefore we have access, through this contribution, to the parton
distributions in the virtual photon.

4) With the ``natural" scale $Q^2+E_\bot^2$, the total cross section is
in good  agreement with the H1 data, thus suggesting that a sizeable BFKL type
contribution may not be necessary to explain the data.

The second point above is important because it does not allow us to make
stable predictions for the cross section and to assess the need for
other contributions of the BFKL-type. Preliminary studies of the
renormalization scale dependence have been performed separately, for
the direct cross section \cite{Aurenche:2003by} and the resolved cross section
\cite{Fontannaz:2004ev}. Here we would like to do a more complete study of the 
scale
sensitivity of the total cross section, including also the effects of
the various factorization scales. As is well known, only the total
cross section has a physical meaning. The separate contributions, Born
terms, HO terms, direct or resolved terms are all factorization and
renormalization scale dependent. Since the direct and resolved NLO
cross section are available, such a study of the scale sensitivity is
now feasible. \par

\begin{figure}
\vspace{9pt}
\begin{center}
\includegraphics[width=4in,height=3.5in]{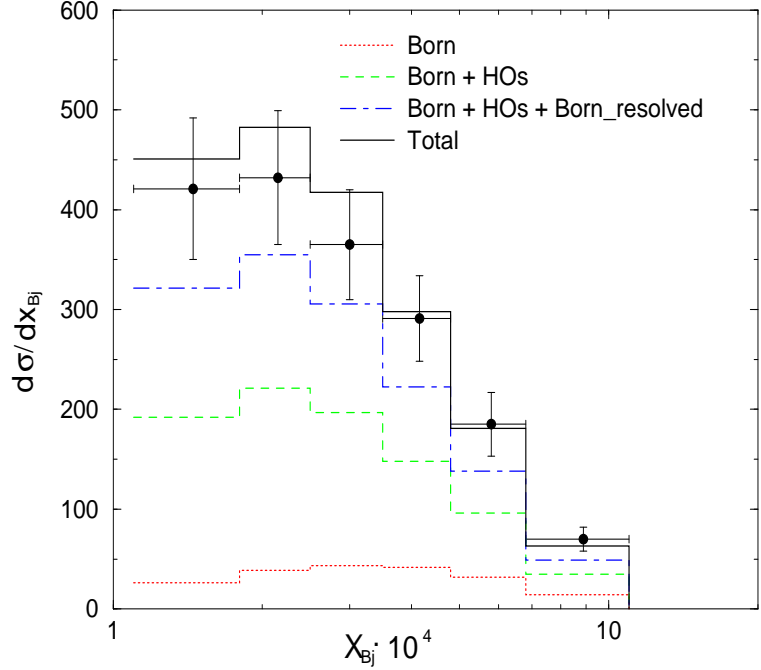}
\end{center}
\caption {The cross section $d\sigma/dx_{Bj}$ corresponding to the range
4.5~GeV$^2 \leq Q^2 \leq 15$~GeV$^2$ and  $E_\bot > 2.5$~GeV
compared to H1 data \cite{h1-2004}. Cuts on all the other kinematical 
variables are given in the text. The symbol $HO_s$ denotes the direct HO
correction from which the lowest order resolved contribution has been subtracted.}
\label{fig:1}
\end{figure}

Before starting this study let us specify the various building blocks of the
total cross section. For the parton distributions in the proton, we use the
CTEQ6M tables~\cite{cteq}, and for the distributions in the virtual photon the
parametrization given in Ref.~\cite{Fontannaz:2004ev}. In the latter
this parametrization was used with fixed value of $Q^2$ corresponding
to the average value $<\!\!Q^2\!\!>$ observed in a cross section. For instance 
for the cross section $d\sigma/dx_{Bj}$ in the range $4.5$~GeV$^2 \leq Q^2 \leq
15$~GeV$^2$ (see Fig.~\ref{fig:1}), we used $<\!\!Q^2\!\!> = 8$~GeV$^2$. This value
corresponds to the overall bin $1.1\ 10^{-4}\leq x_{Bj} \leq 11.0\ 10^{-4}$.
However this value changes with $x_{Bj}$ and the description of the whole
$x_{Bj}$ domain by a single value $<\!\!Q^2\!\!>$ is not accurate. Therefore in this
paper we use a parametrization depending continuously on $x_{Bj}$, $Q^2$ and
$M_{\gamma}^2$. We work in the $\overline{\rm MS}$ renormalization and
factorization schemes and all the scales are equal to  $(Q^2 + E_{\bot}^2)$. We
take $n_f = 4$ flavors and for $\alpha_s(\mu )$ we use an exact solution of the
two-loop renormalization group equation with $\Lambda_{\overline{MS}} =
326$~MeV. The fragmentation functions of the partons in $\pi^0$ are those of
Ref. \cite{kkp}. With these inputs we obtain the cross section displayed in
Fig.~\ref{fig:1} and compared with  H1 data measured in the range 4.5~GeV$^2 
\leq Q^2 \leq 15$~GeV$^2$ \cite{h1-2004}. Our calculations are performed at
$\sqrt{S} = 300.3$~GeV and the forward-$\pi^0$ cross section is defined with
the following cuts. In the laboratory system a $\pi^0$ is observed in the
forward direction with $5^{\circ} \leq \theta_{\pi} \leq 25^{\circ}$~; the
laboratory momentum of the pion is constrained by $x_{\pi} =
E^{lab}_{\pi}/E^{lab}_p \geq 0.1$, and an extra cut is put on the $\pi^0$
transverse momentum in the $\gamma^*-p$ center of mass system:  $E_{\bot}
>2.5$~GeV. The inelasticity $y = Q^2/x_{Bj}S$ is restricted to the range 
$0.1 < y < 0.6$.
\vskip 3 truemm
We clearly observe in this figure the points 1), 3) and 4) mentioned
above, and in particular, the very large HO$_s$ (the index $s$ means that
the lowest order resolved component has been subtracted from the HO
corrections to the direct term, as discussed at the beginning of this
section). We also notice the importance of the HO corrections to
the resolved cross section. With respect to the corresponding figure of
Ref. \cite{Fontannaz:2004ev}, we note that the resolved component is larger at 
small
$x_{Bj}$ and smaller at large $x_{Bj}$, which improves agreement with data at
large $x_{Bj}$. This is due to the fact that
the average $<\!\!Q^2\!\!>$ is smaller at small $x_{Bj}$ than at large
$x_{Bj}$. 
Figure~\ref{fig:1} is the starting point of our scale studies. We choose a
kinematic region  for which the HO corrections are large (this is due to the
small value of $E_{\bot} > 2.5$~GeV) in order to better
exhibit the scale dependence, but with the consequence (as we shall
see) that the cross
section does not stabilize for an optimum choice of the scales.
Let us define the factorization scales $M_k^2 = C_k^2 (Q^2+E_\bot^2)$
where $k$ stands for I
(the proton distribution scale) or F (the fragmentation function (FF) scale).
We also introduce the virtual photon factorization scale
$M_\gamma^2=Q^2+C_\gamma^2 E_\bot^2$ and
the renormalization scale $\mu^2=C_\mu^2 (Q^2+E_\bot^2)$.

\begin{figure}
\vspace{9pt}
\begin{center}
\includegraphics[width=2.in,height=2.5in]{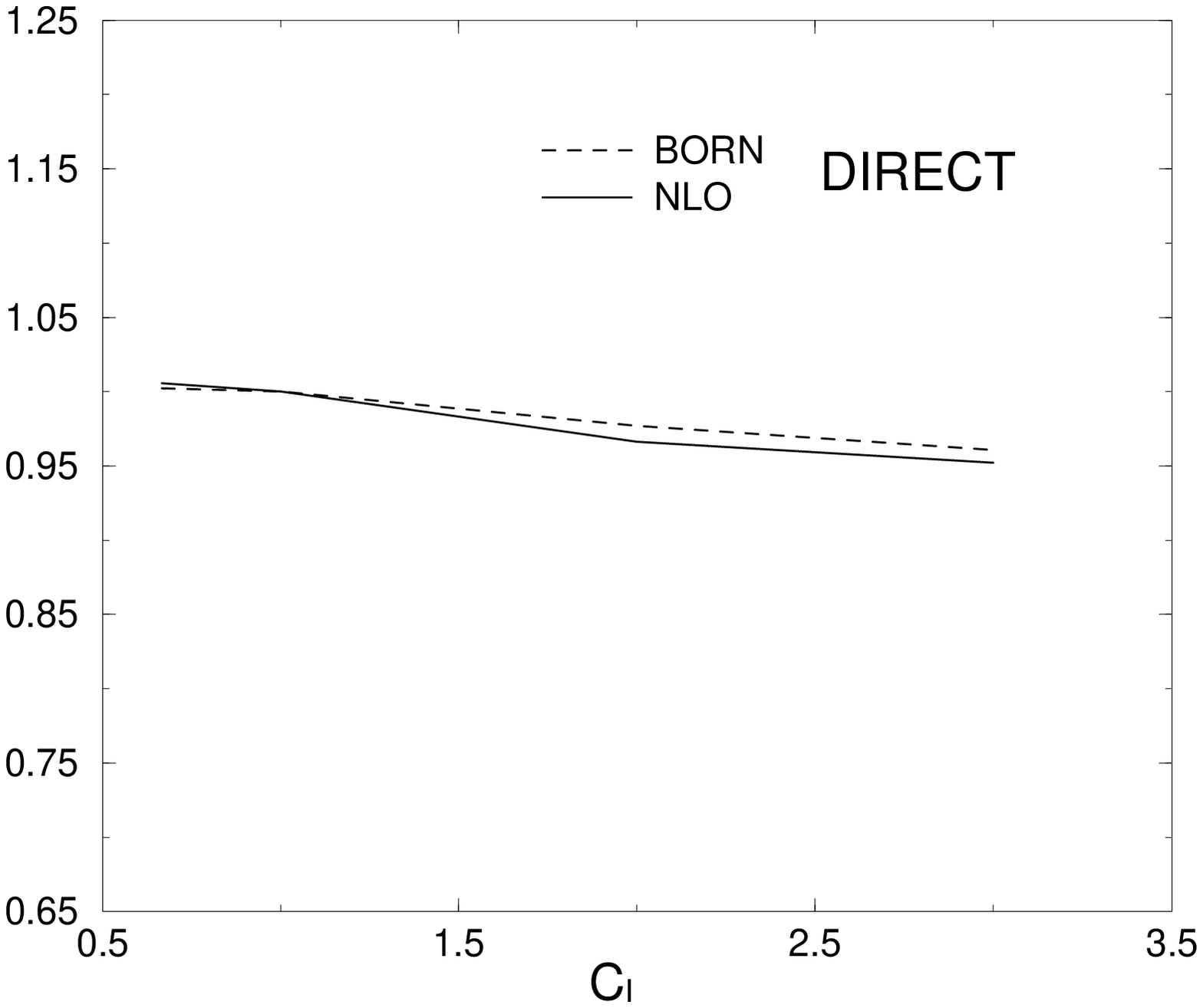}
\includegraphics[width=2.in,height=2.5in]{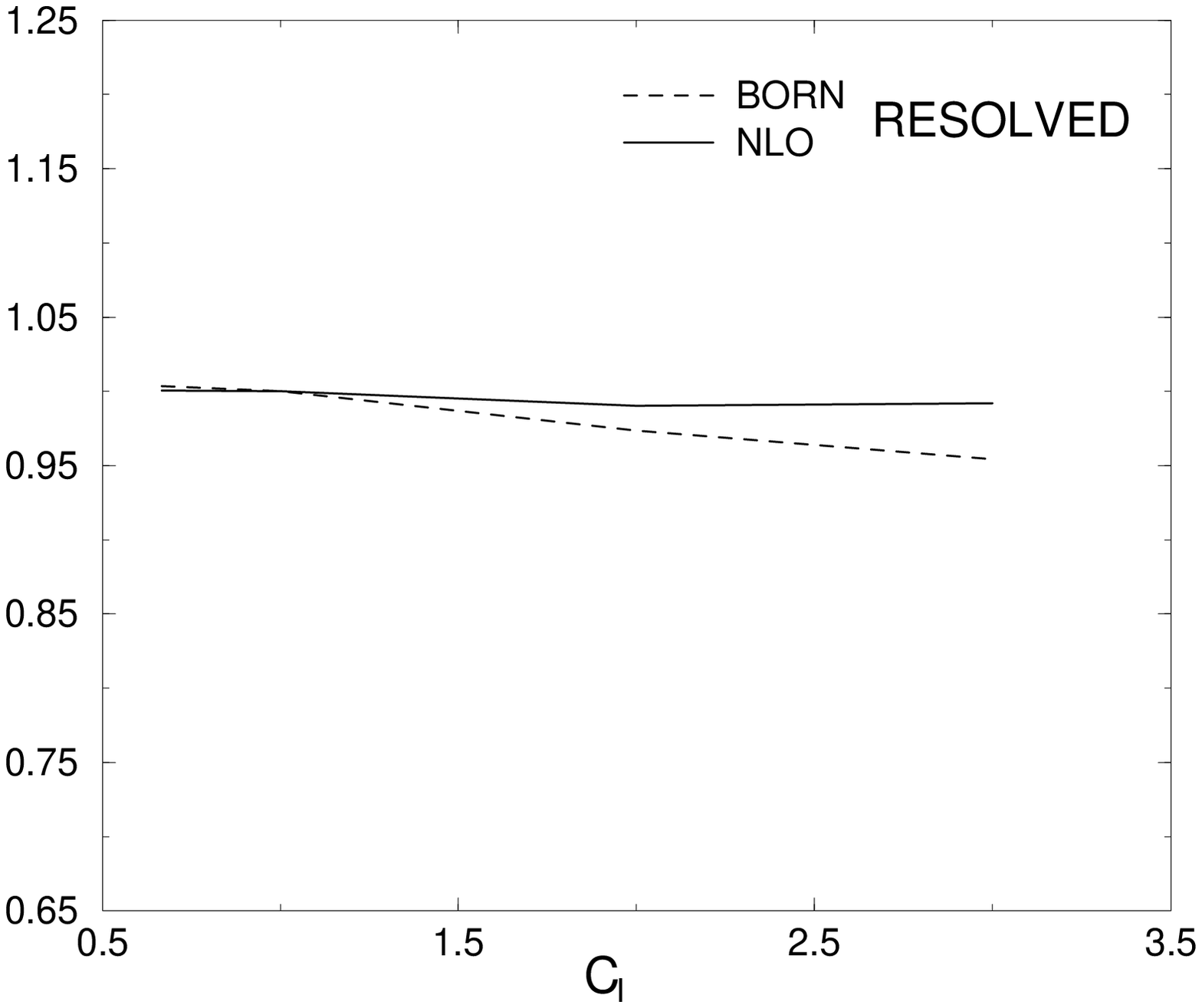}
\includegraphics[width=2.in,height=2.5in]{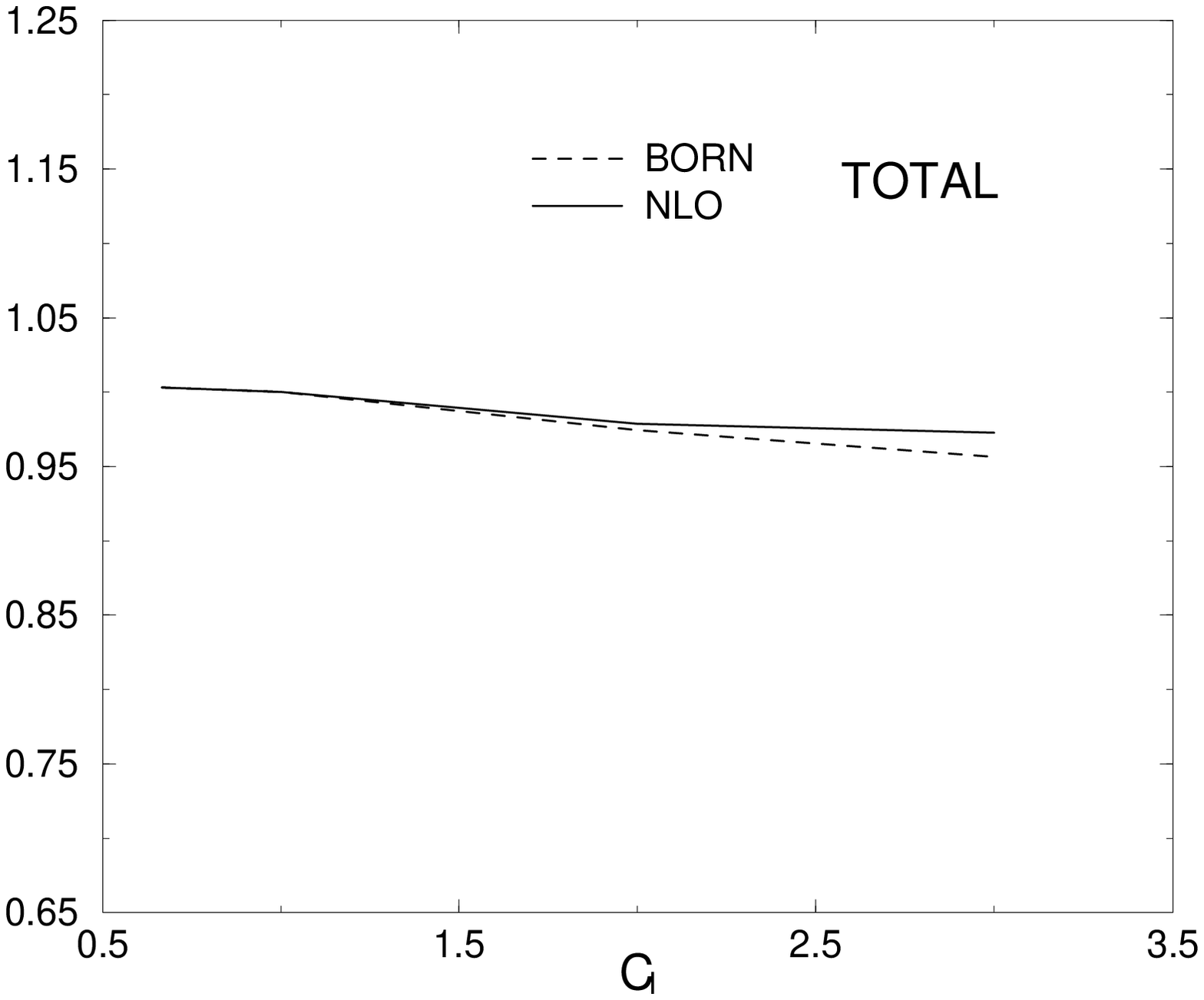}
\end{center}
\caption {Cross section variations with $C_I$ which  has been  defined in 
the text. The cross sections are normalized to 1.0 at $C_I=1.$ The choice of 
kinematic conditions is the same as in  Fig.1.}
\label{fig:2}
\end{figure}
\vskip 3 truemm
\begin{figure}
\vspace{9pt}
\begin{center}
\includegraphics[width=2.in,height=2.5in]{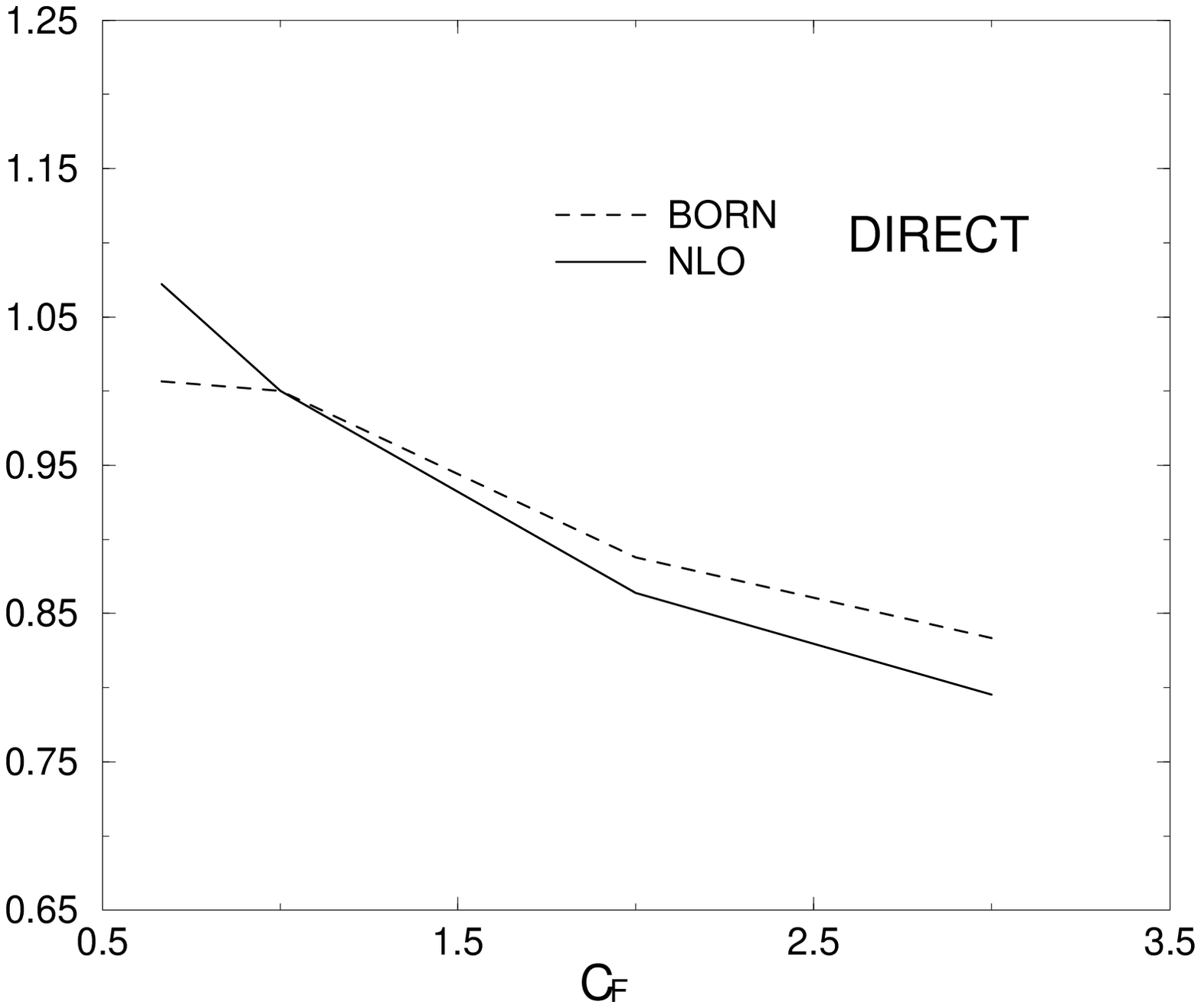}
\includegraphics[width=2.in,height=2.5in]{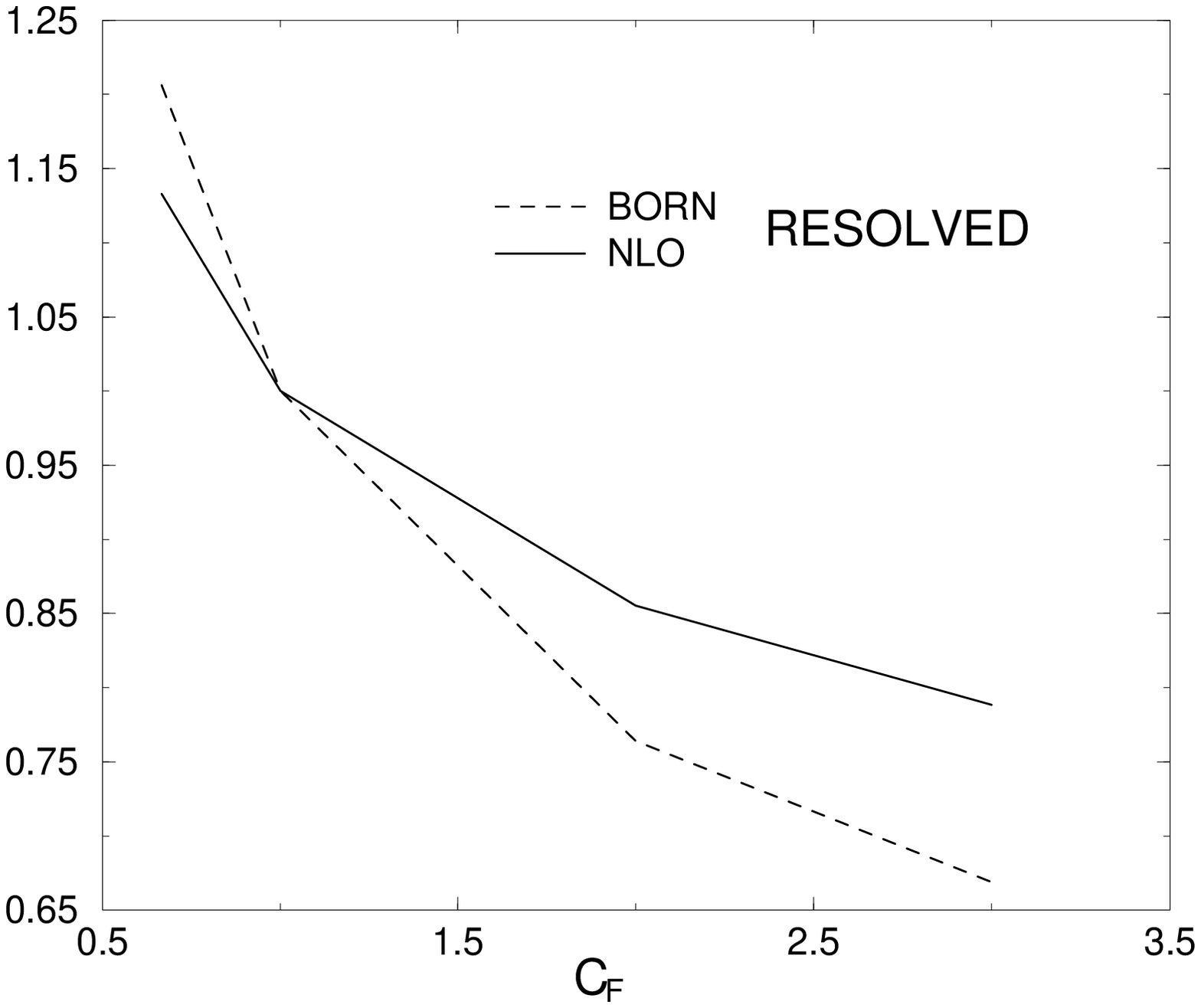}
\includegraphics[width=2.in,height=2.5in]{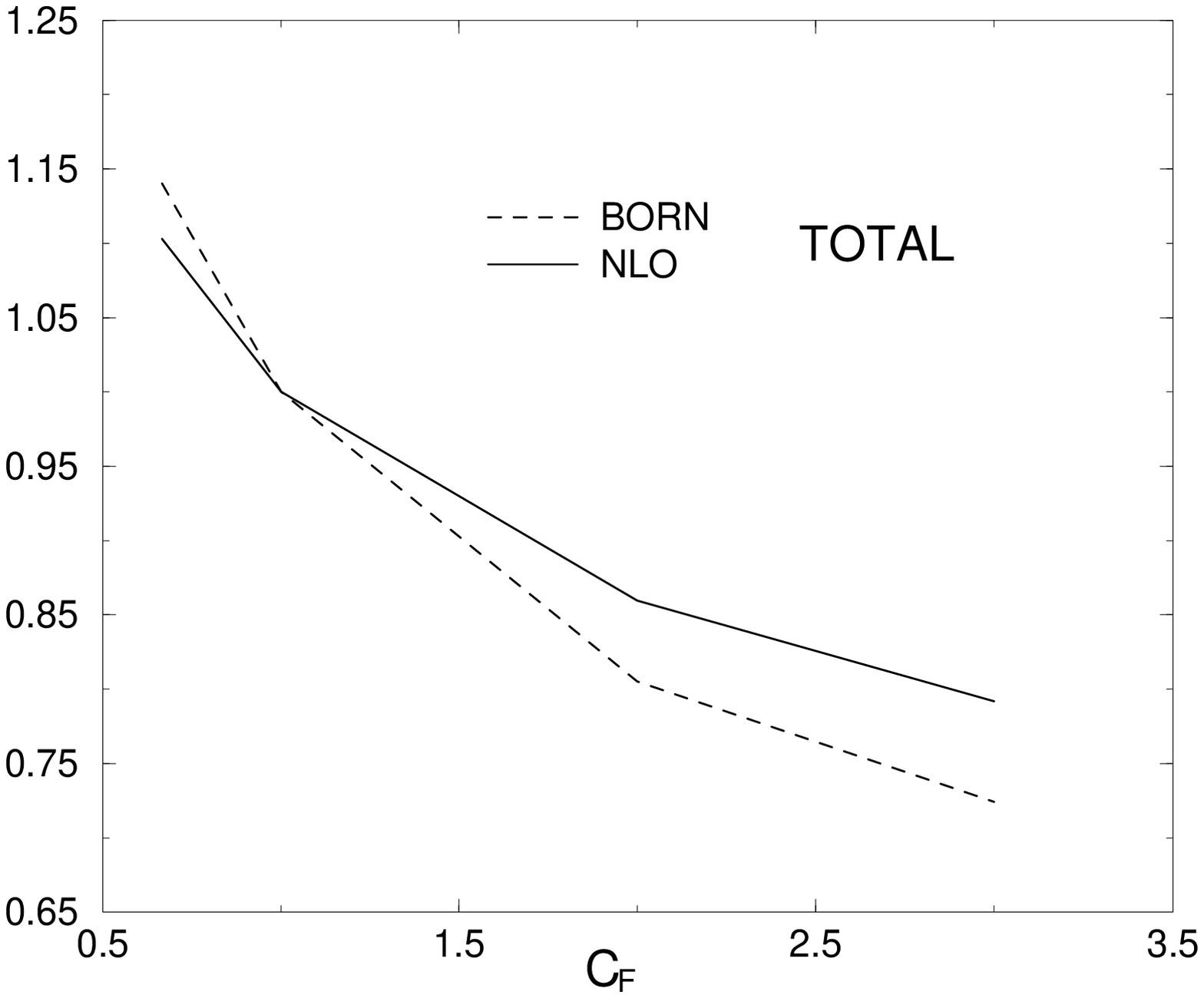}
\end{center}
\caption {Cross section variations with $C_F$ which has been defined in the
text. The cross sections are normalized
to 1.0 at $C_F=1.$ The choice of kinematic conditions is the same
as in Fig.1.}
\label{fig:3}
\end{figure}
\vskip 3 truemm
We study the sensitivity of the various components of Fig.~1 in the single bin
$1.1\times 10^{-4} < x_{Bj} < 11.0\times 10^{-4}$ and start with the 
factorization scales $M_I$ and $M_F$. In Figs.~2 and 3, we observe very
different behavior. The variation with $M_I$ is almost flat whereas that with
$M_F$ is strongly decreasing. These differences are due to the different average
values of $x_p$, the proton distribution variable, and $z$, the fragmentation
function variable, corresponding to the kinematics of Fig.~1. In the direct 
process for instance, we have $<\!\!x_p\!\!> \sim 0.1$, a domain in which the 
proton distribution functions do not vary much with $M_I$. For the 
fragmentation 
variable we have $<\!\!z\!\!> \sim 0.3$ in the direct case, and $<\!\!z\!\!>
\sim 0.7$ in the  resolved case. In these ranges, the variation of the
fragmentation functions $D(z,M_F)$ are not  negligible; the higher the value of
$z$, the stronger the variation. Hence the different behavior of the direct and
resolved contributions. Moreover in the  direct case, we observe that the HO
corrections do not compensate the Born term variation. This is due to the fact
that the HO corrections contain new  channels which appear as new Born
contributions for which there are no compensating $\ln(M_F/E_\bot)$ terms. For
instance, we have the opening of the  new 
\begin{figure}
\vspace{9pt}
\begin{center}
\includegraphics[width=3.5in,height=2.5in]{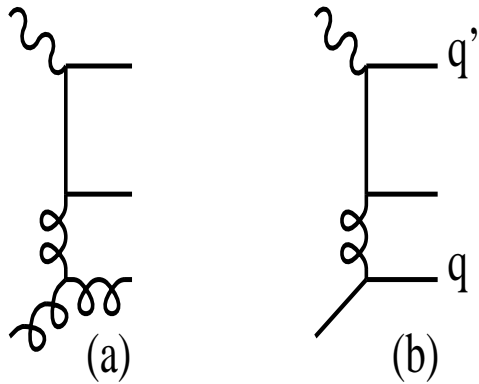}
\end{center}
\caption {Examples of HO graphs leading to the opening of new channels when 
the final hadron is a fragment of the gluon or of the quark q.}
\label{fig:4}
\end{figure}
channel corresponding to Fig.~4a with the final gluon fragmenting into a
$\pi^0$. At
NLO there is no counter term which corrects this new Born contribution. Such
terms would only appear at NNLO. The contribution of this new channel, 
involving the exchange of a gluon in the $t$-channel, is very large and the 
overall behavior of the NLO cross section is very similar to that of the Born
contribution. In the resolved contribution a graph with a gluon exchanged in 
the $t$-channel already exists at the Born level and the HO corrections contain
the appropriate counter term. However the compensation between the Born 
contribution and the HO is not complete due to the large values 
of $<\!\!z\!\!>$ involved.
\begin{figure}
\vspace{9pt}
\begin{center}
\includegraphics[width=2.in,height=2.5in]{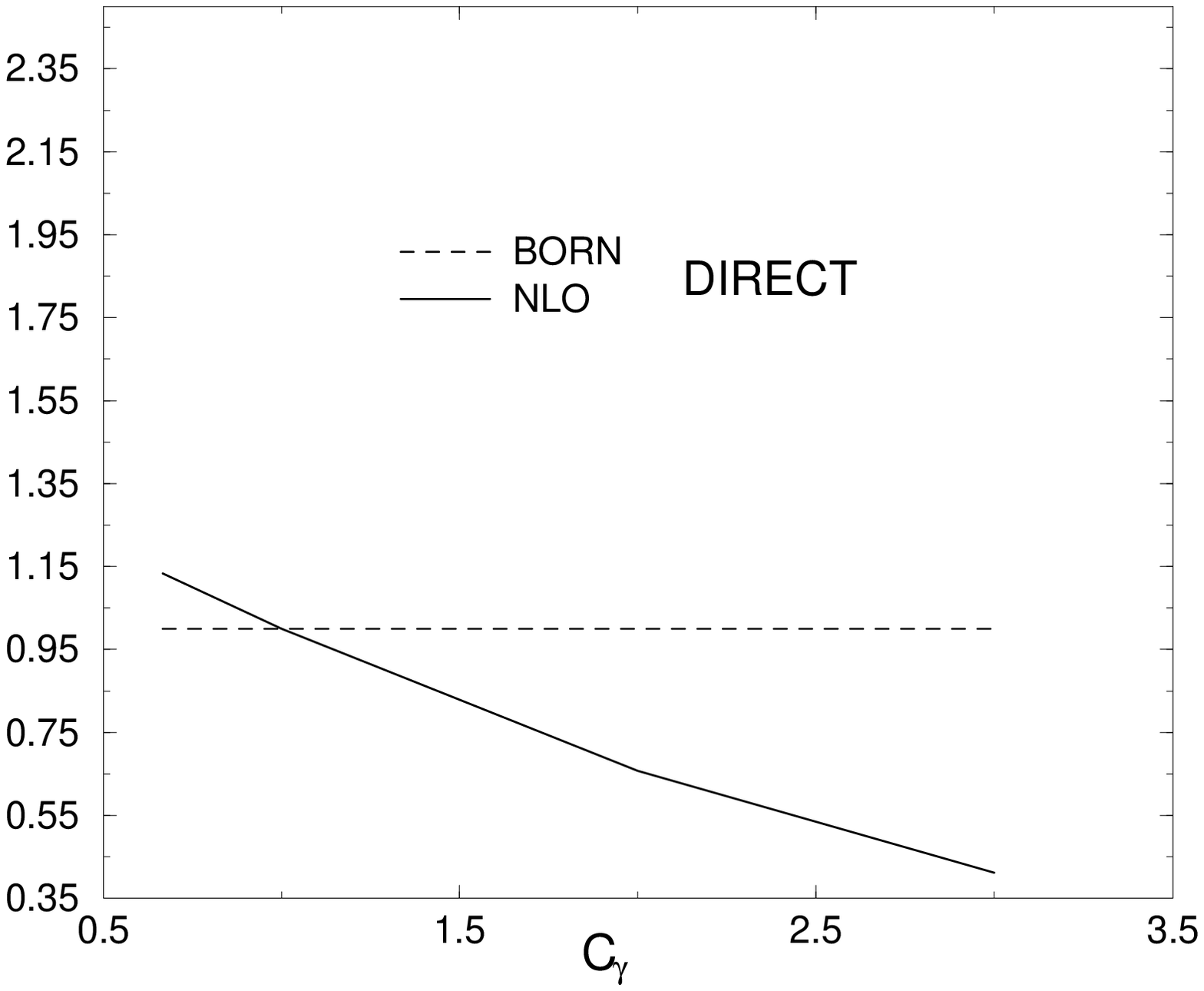}
\includegraphics[width=2.in,height=2.5in]{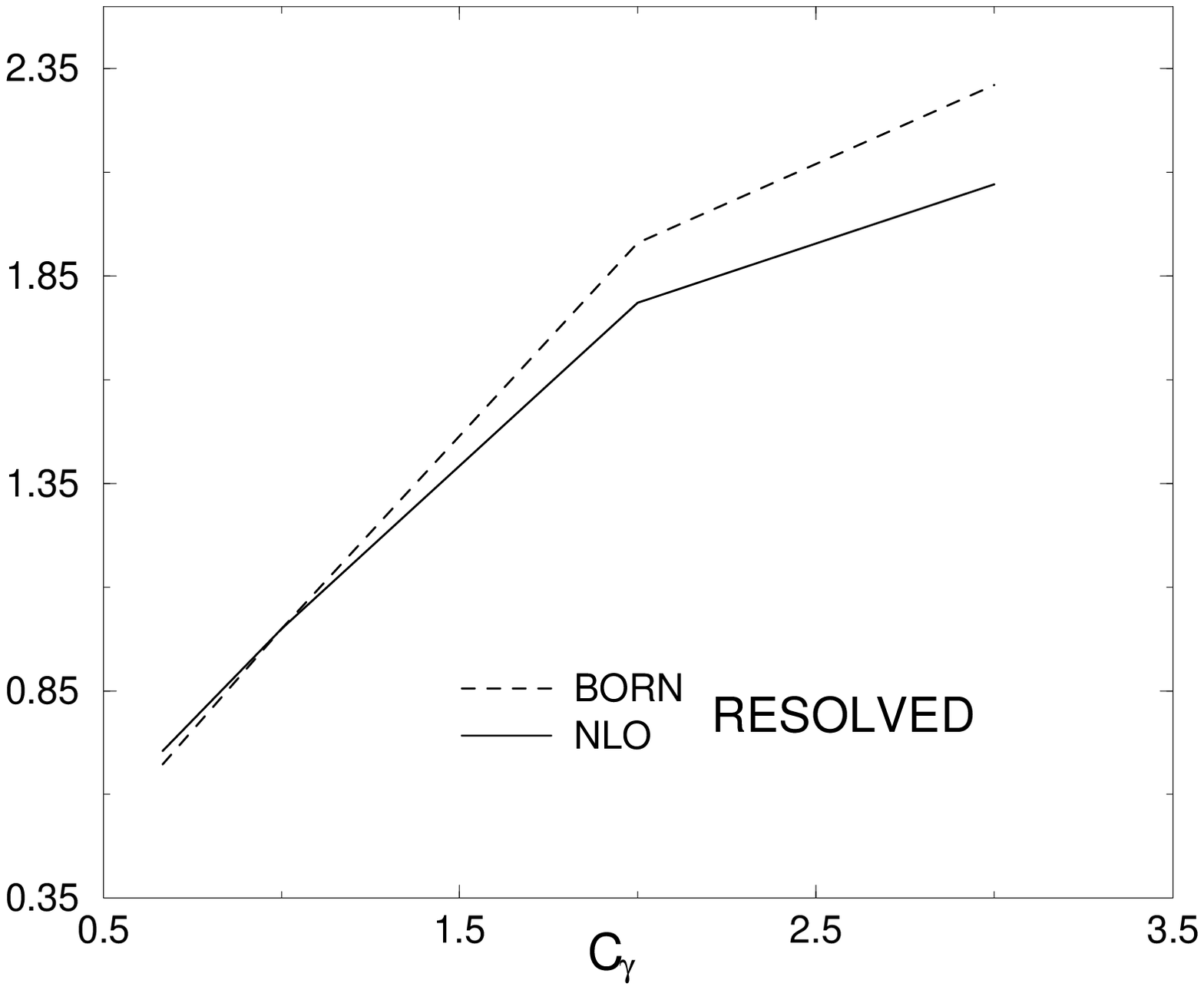}
\includegraphics[width=2.in,height=2.5in]{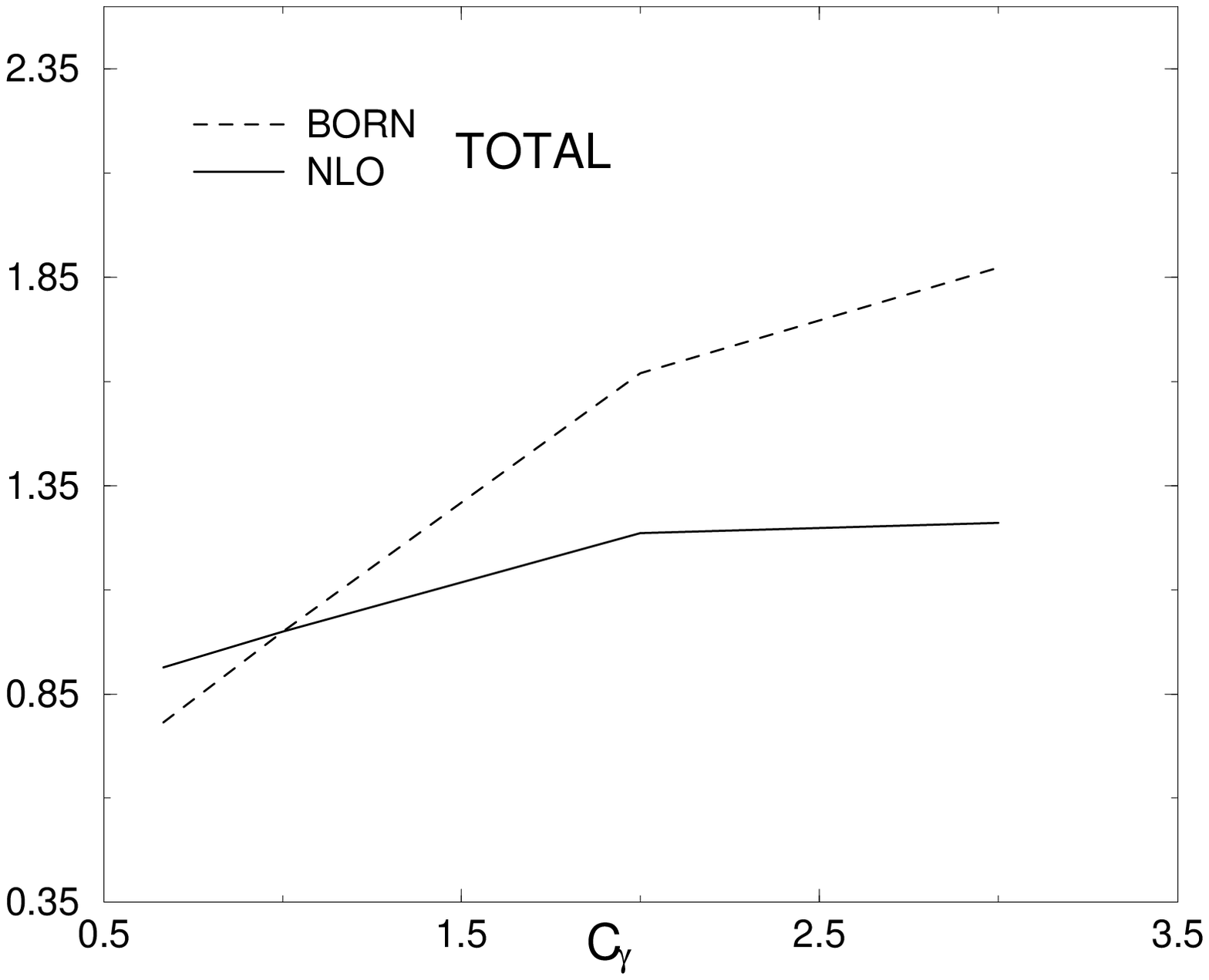}
\end{center}
\caption {Cross section variations with $C_\gamma$ which has been defined in
the text. The cross sections are 
normalized to 1.0 at $C_\gamma=1.$ The choice of kinematic conditions is 
the same as  in Fig.1.}
\label{fig:5}
\end{figure}

Let us now turn to the study of the variations with $M_\gamma$ which are 
displayed in Fig.~5. In the variations studied till now, there was no
compensation between the resolved and the direct terms. For instance, the $M_I$
dependence  of the parton distributions in the photon were separately
compensated by $\ln(M_I/E_\bot)$ terms which appear in the direct or in the
resolved HO corrections. However, for $M_\gamma$, we have compensation
between the resolved and  the direct terms that we can observe in Fig.~5. When
$M_\gamma$ increases the Born+ HO$_s$ direct contribution decreases. This is
due to the fact that a term proportional to $\log(M_\gamma/Q)$ is subtracted
from the HO corrections leaving a piece $\log(E_\bot/M_\gamma)$ in the
remaining HO$_s$ part, as explained at the beginning of this section. On the
other hand, the resolved  contribution increases with the increase of the parton
distributions in the virtual photon. A counter term present in the HO resolved
correction dampens  the variation of the NLO cross section compared to the Born
case. More precisely, the scale variation of the photon structure function
contains two pieces (see {\it i.e.} eq.~(16) in Ref.~\cite{Fontannaz:2004ev}):
the inhomogeneous part, proportional to $\alpha$, and the homogeneous or
hadron-like part proportional to $\alpha \alpha_s$. The scale variation of the
inhomogeneous part is compensated by the HO$_s$ direct term, while that of the
homogeneous part is in the HO resolved contribution. For a consistent
calculation it is therefore necessary to work at the NLO level for both direct
and resolved pieces. Due to the compensation between the direct and the 
resolved contributions, the total cross section exhibits a smoother behavior 
when $M_\gamma^2$ varies by a  factor 20.

Finally let us consider the variations as a function of the renormalization
scale $\mu$. 
\begin{figure}
\vspace{9pt}
\begin{center}
\includegraphics[width=2.in,height=2.5in]{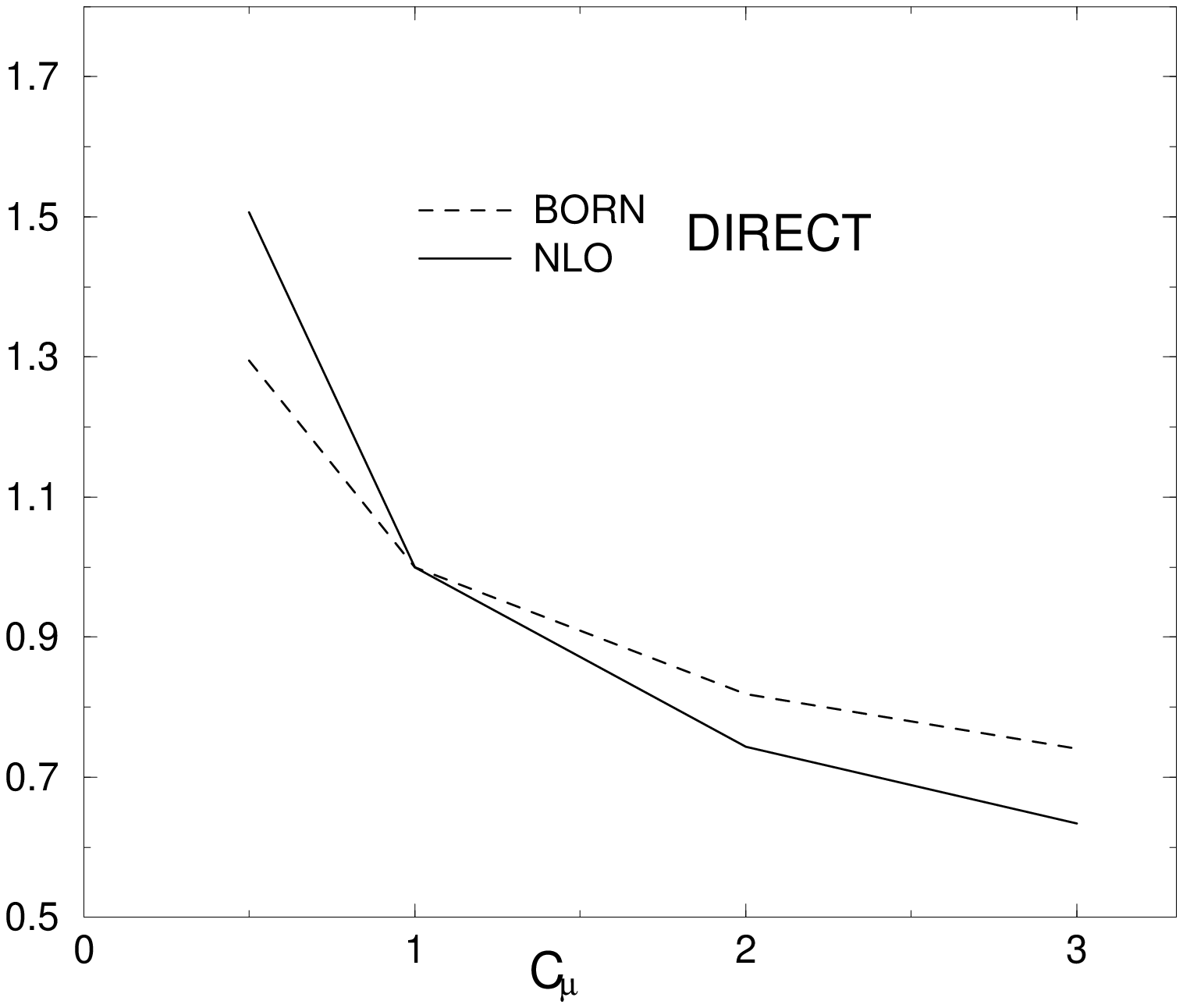}
\includegraphics[width=2.in,height=2.5in]{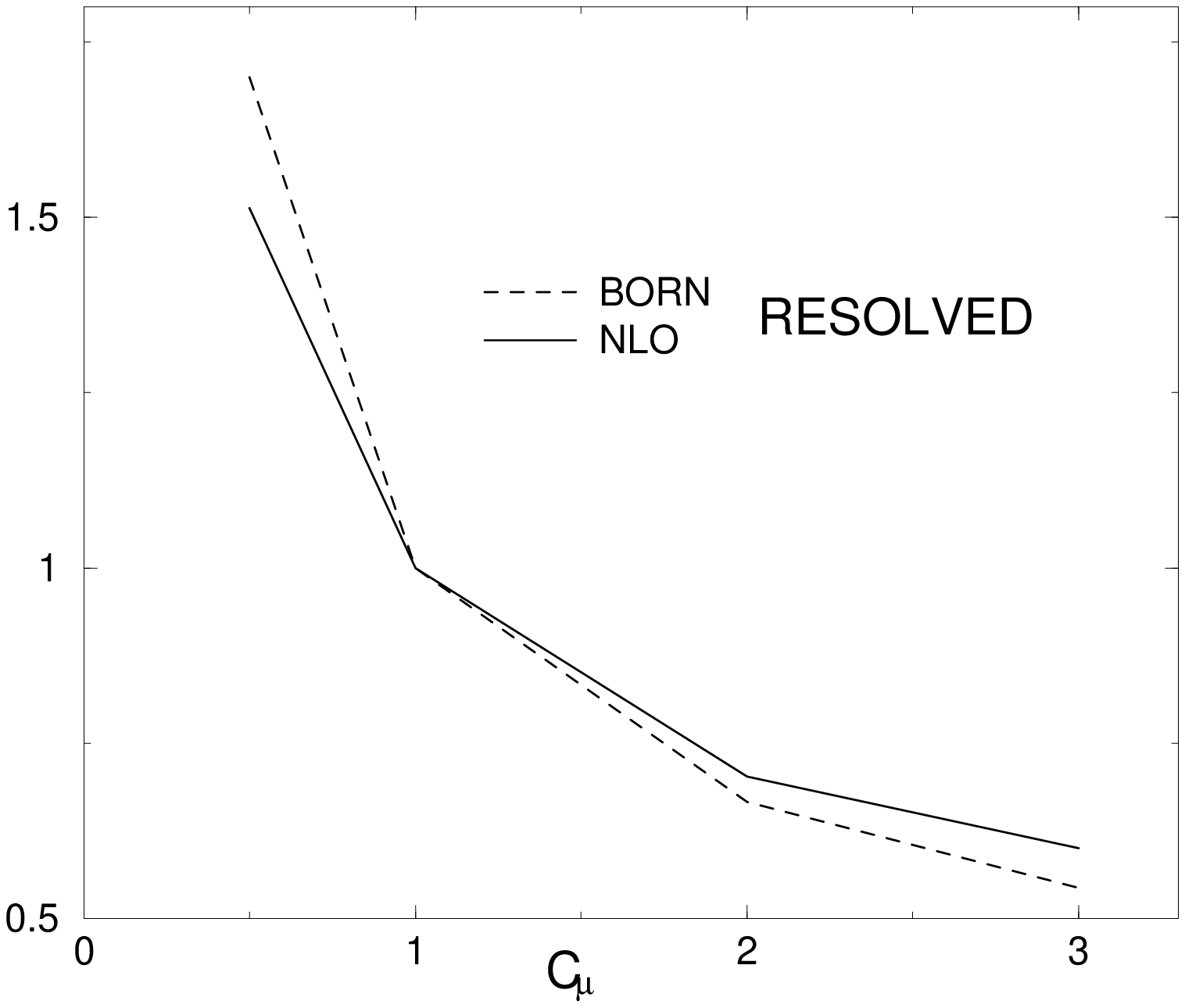}
\includegraphics[width=2.in,height=2.5in]{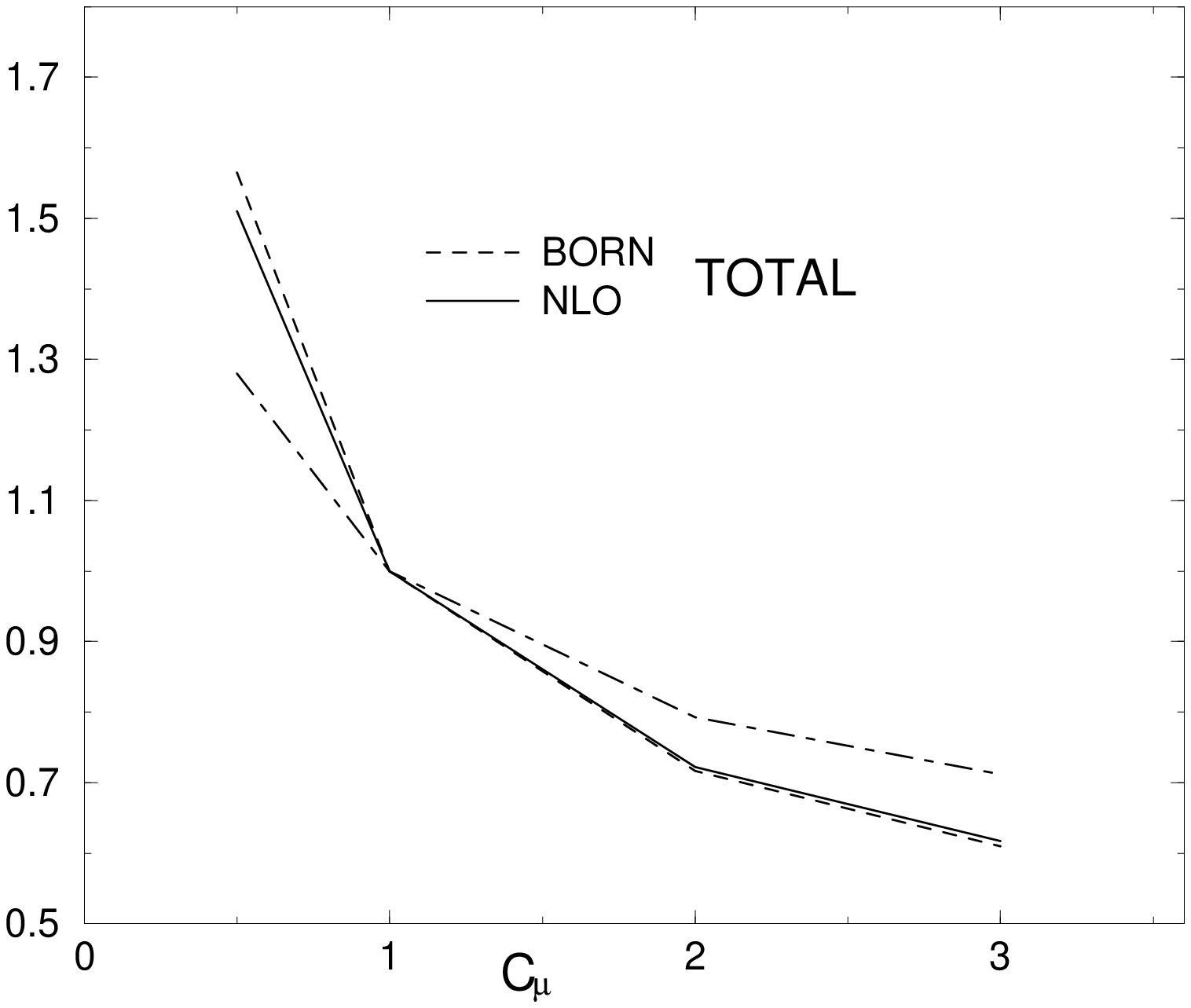}
\end{center}
\caption {Cross section variations with $C_\mu$ which has been defined in 
the text. The cross sections are 
normalized to one at $C_\mu=1.$ The choice of kinematic conditions is the same
as in Fig.1 except for the rightmost panel, where the dash-dotted  curve
corresponds to the NLO cross section calculated in the bin  $20~{\rm GeV}^2 <
Q^2 < 70~{\rm GeV}^2$, $3.9 \times 10^{-4} < x_{B_j} < 6.3  \times 10^{-3}$,
$E_\bot > 3.5$ GeV.}
\label{fig:6}
\end{figure}
They are the largest. We note the same phenomenon as observed
for the $M_F$-scale variation: no compensation for the direct NLO term and a
small compensation for the NLO resolved contribution. Concerning the direct 
term, this behaviour again arises due to the opening of new channels, without
virtual corrections (they appear only at NNLO), containing terms in
$\log(\mu/E_\bot)$ to compensate the $\mu$ dependence of $\alpha_s(\mu)$. 
As these new Born terms are
proportional to $\alpha_s^2(\mu)$ and constitute a large part of the cross 
section \cite{Aurenche:2003by},
the variation of the latter is strong. On the other hand virtual corrections
to terms containing a gluon exchanged in the t-channel are present in the
resolved contribution. This produces the small effect observed in Fig.~5 and we
do not find any reasonable value of $\mu$ for which the cross section would 
reach an optimum. This is due to the large HO corrections corresponding to the 
small values of the transverse energy in the H1 kinematical domain studied here.
Indeed the H1 experiment puts a minimum cut-off on $E_\bot$ (the transverse
energy in the $\gamma^*$-proton center of mass frame) of 2.5 GeV. This is a 
small  value for a ``large-$p_T$'' reaction and the resulting HO are large. 
However for higher values of the cut-off, the HO corrections are smaller and 
we find a $\mu$-variation of the resolved cross section which exhibits an 
optimum (maximum) point. For instance in Ref.~\cite{Fontannaz:2004ev}  
a cut-off $E_\bot > 5$~GeV  was used and an optimum of the cross section 
was found for $C_\mu \sim 0.2$.

Therefore we reach the conclusion that the addition of the NLO resolved 
component improves 
the behavior of the cross section with respect to the scale variation.
However the sensitivity of the cross section
to the renormalization scale variation prevents us from predicting absolute 
values for
the latter. For instance, in the range $1/4<C^2_\mu<4$, the predictions vary by 
a factor 2. This fact clearly points towards the necessity of calculating NNLO
corrections. For the time being with the aim of phenomenological applications
in mind, we choose scales which lead to a good description of the data in the
range 4.5~GeV$^2  \leq Q^2 \leq 15$~GeV$^2$. As we can see from Fig.~1, such an
agreement is found with all scales set equal to $Q^2+E^2_{\bot}$. Then with the
same scales we make predictions for $d \sigma/dx_{Bj}$ in the 
other $Q^2$-ranges,
as well as for $d \sigma/dE_\bot$ and $d \sigma/dx_{\pi}$. Because of the 
marked scale sensitivity
of the cross sections, scales giving a good description of data in a given 
$Q^2$-range do not necessarily lead to a good agreement in another range. 
It turns 
out, as we shall see, that a satisfactory description of all the data can be 
obtained with this single choice of scales. Of course the scale choice could 
be refined in order to improve the agreement between data and theory in Fig.~1, 
especially at small $x_{Bj}$ . But this is a formal exercise that does not 
present any physical interest.

Finally, to ameliorate some of our negative conclusions on the scale 
dependence of the cross
section, we note that at larger $Q^2$ and $E^2_{\bot}$, the sensitivity to the
renormalization scale is reduced. In Fig.~6 (rightmost panel), we display the 
behavior of the total NLO cross section in the range 
20~GeV$^2  \leq Q^2 \leq 70$~GeV$^2$ with 
$3.9\times 10^{-4} < x_{Bj} < 6.3\times 10^{-3}$
($E_{\bot} > 3.5$~GeV). The cross section varies by less than $\pm$~25\% when
$C_\mu$ is in the range $1/4 <C^2_\mu< 4.$
\begin{figure}[!tb]
\vskip 1cm
\begin{center}
\hskip -2.cm
\includegraphics[width=15cm]{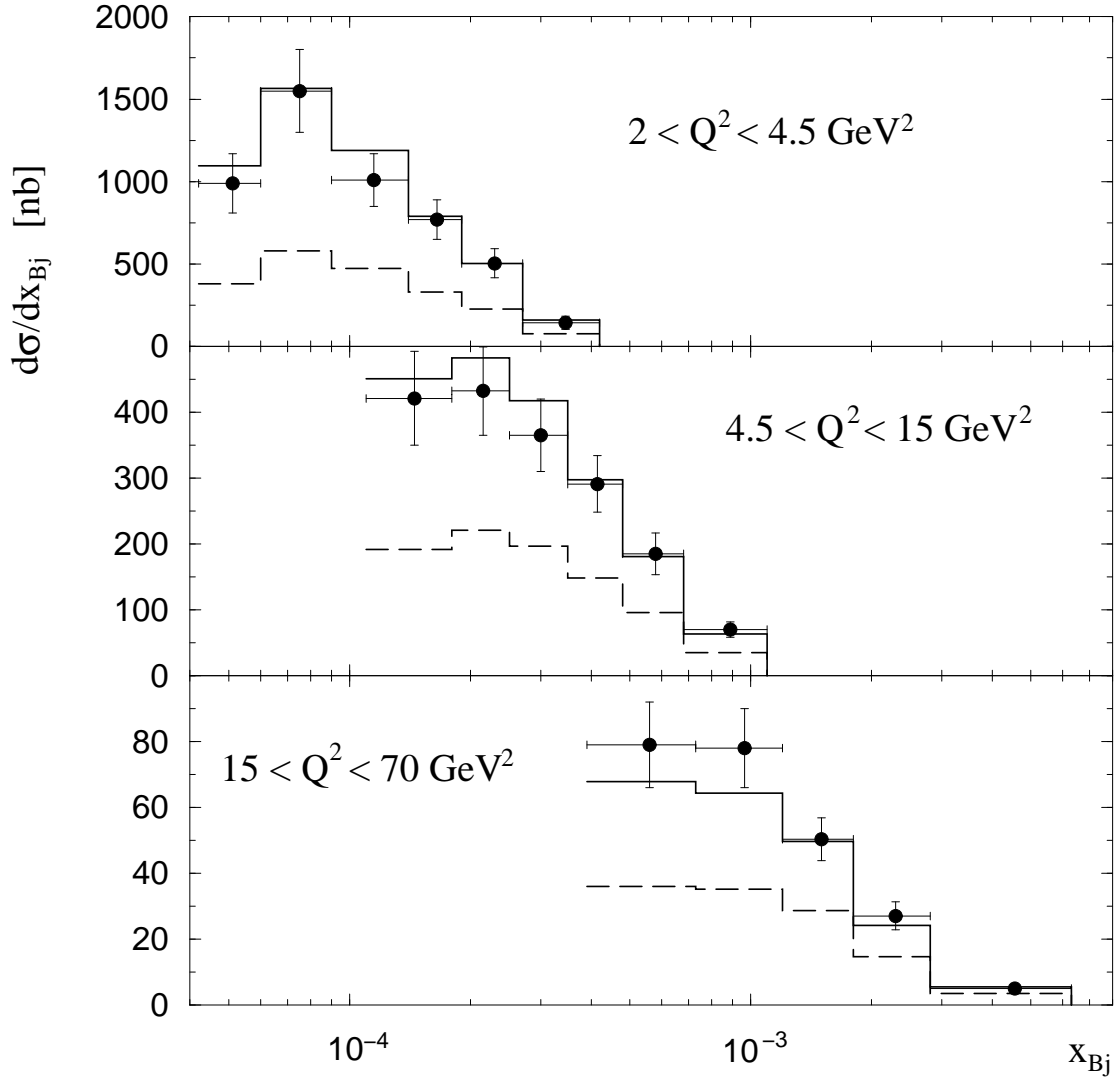}
\end{center}
\caption {
Inclusive $\pi^0$ cross section as a function of $x_{Bj}$ in  the range $E_\bot
>$ 2.5 GeV for three different intervals in $Q^2$. The cuts on other variables
are given in the text. The data points are from the H1
collaboration~\cite{h1-2004}. The histograms are the theoretical predictions
obtained with all scales set equal to $(Q^2+E_\bot^2)$; solid line: full NLO
predictions; dashed line: ``direct" contribution.}
\label{fig-xbj-25}
\end{figure}

\section{Comparison to H1 data}

We are now ready to compare the theoretical predictions to the H1 recent
results~\cite{h1-2004} on single $\pi^0$ inclusive cross section. The same
kinematical cuts as in the experimental data are imposed on the theory, and they
are given in the previous section while discussing Fig.~\ref{fig:1}.

Concerning the theoretical predictions ``NLO'' will refer to the full
next-to-leading logarithmic predictions for the direct term as well as for the
resolved term, where ``direct'' refers to the lowest order (``Born'') term with
the attached higher order corrections labeled ``HO${}_s$'' above\footnote{In
\cite{Kniehl:2004hf} an  extensive discussion is given of the interference
terms where the photon couples to two different quark lines, which leads to
triangle graphs when calculating the cross section (the so-called 
Furry terms) and these terms are found to give an appreciable contribution. In
the present calculation, valid for neutral pion production, these terms are not
present because the quark production cross section is cancelled by the antiquark
cross section. The same reason makes them vanish in jet production.}. The
$\overline{\rm MS}$ scheme is used throughout with $\Lambda_{\overline{MS}} =
326$~MeV. For convenience we recall here the basic ingredients entering the
calculation. All predictions are made using CTEQ6M~\cite{cteq} for the proton
parton distributions and KKP~\cite{kkp} for the fragmentation functions of the
pion. For the virtual photon structure function, in the resolved term, the
recent parametrization of Ref.~\cite{Fontannaz:2004ev} is taken.  Using the
lowest order approximation (the so-called ``box" approximation) changes the
results by less than 10\%. The common scale is chosen to be  $(Q^2+E_\bot^2)$.

\begin{figure}[!tb]
\begin{center}
\hskip -2.cm
\includegraphics[width=15cm]{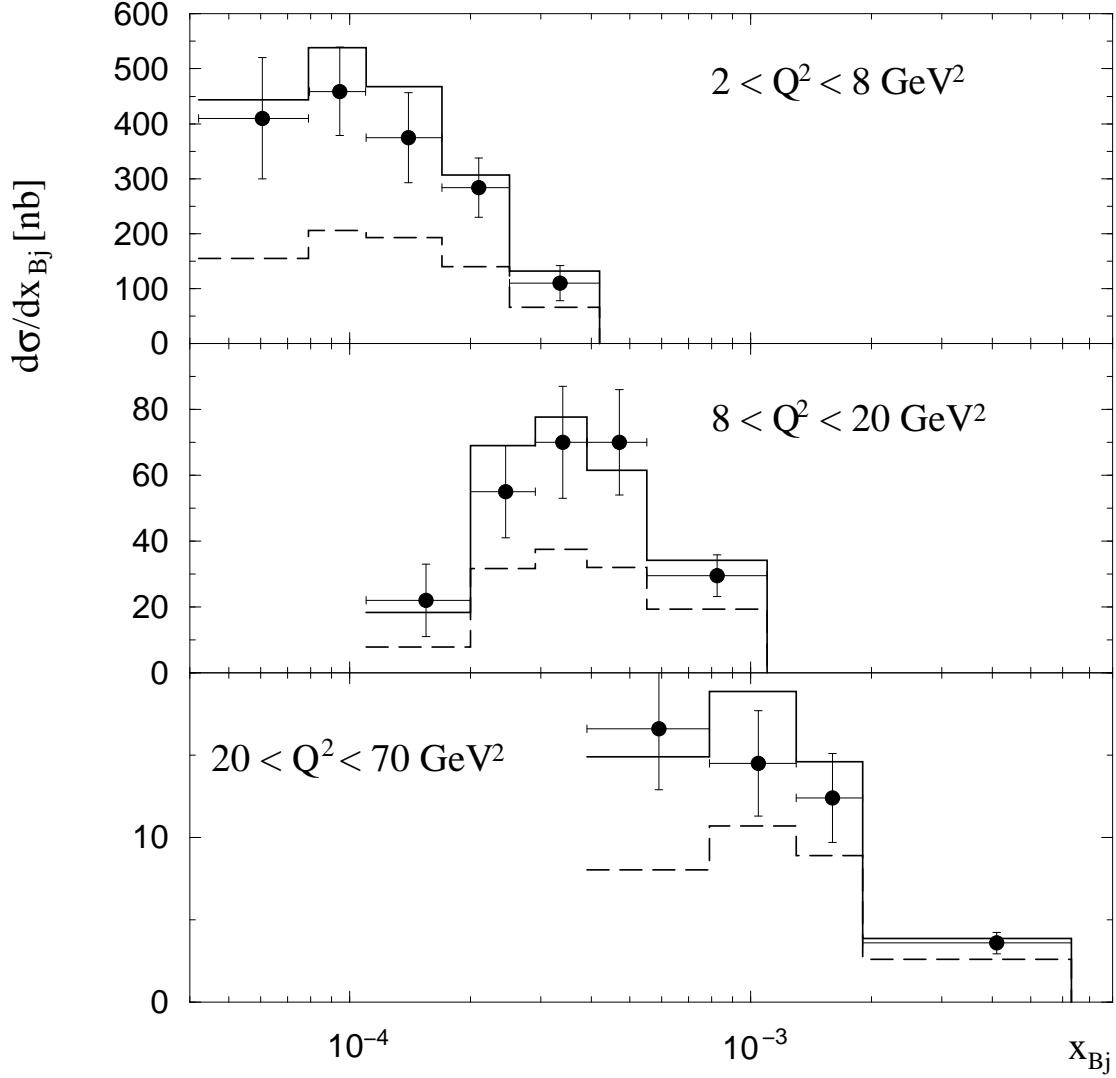}
\end{center}
\caption {Inclusive $\pi^0$ cross section as a function of $x_{Bj}$ in the
range $E_\bot >$ 3.5 GeV for three different intervals in $Q^2$. The cuts on
other variables are given in the text. The data points are from the H1
collaboration~\cite{h1-2004}. The histograms are the theoretical results: the
solid line corresponds to full NLO predictions and the  dashed line
to the ``direct" contribution. Choice of scales is as in Fig.~\ref{fig-xbj-25}.}
\label{fig-xbj-35}
\end{figure}

The comparison between theory and experiment for the single pion distribution
as a function of $x_{Bj}$ is shown in Figs.~\ref{fig-xbj-25} and
\ref{fig-xbj-35} for the cuts $E_\bot >$ 2.5 GeV and $E_\bot >$ 3.5 GeV
respectively. We notice the very good overall agreement between data and theory
(note the linear scale) for the whole $x_{Bj}$ range. At a finer level one may 
observe some systematics
in Fig.~\ref{fig-xbj-25} where at low $x_{Bj}$, for the medium $Q^2$ range, the
theoretical predictions fall slightly above the data while in the large $Q^2$
bin it is the opposite.  Furthermore, one notes the importance of the resolved
contribution (the difference between the solid and the dashed line). At low
$Q^2$ (upper panels) it is 1.5 to 1.9 times the direct contribution, decreasing
as $x_{Bj}$ increases, while at large $Q^2$ (lower panels) it never exceeds the
direct term and becomes almost negligible at large $x_{Bj}$: this is as
expected from the discussion in the previous section. The importance of the
resolved contribution to obtain agreement with the data was also pointed out by
Kramer and P\"otter who calculated the NLO cross section (resolved term at
leading order) to forward dijet production~\cite{kramer:99} and compared it
with H1~\cite{h1<1999} and ZEUS~\cite{Breitweg:1998ed} data,  as well  
by Jung and collaborators\cite{Jung:1999eb} in their analysis using a LO 
calculation.
One may comment again on
the rather unusual situation at low $x_{Bj}$ where the HO$_s$ correction to the
direct term can be up to an order of magnitude larger than the Born term (see
Fig.~\ref{fig:1}) due to the appearance of the BFKL-like terms of
Fig.~\ref{fig:4} with gluon poles. At large $x_{Bj}$ however one recovers the
``usual'' situation where the HO$_s$ piece is of the same order of magnitude as
the lowest order term. 

\begin{figure}[!htb]
\begin{center}
\hskip -2.cm
\includegraphics[width=15cm]{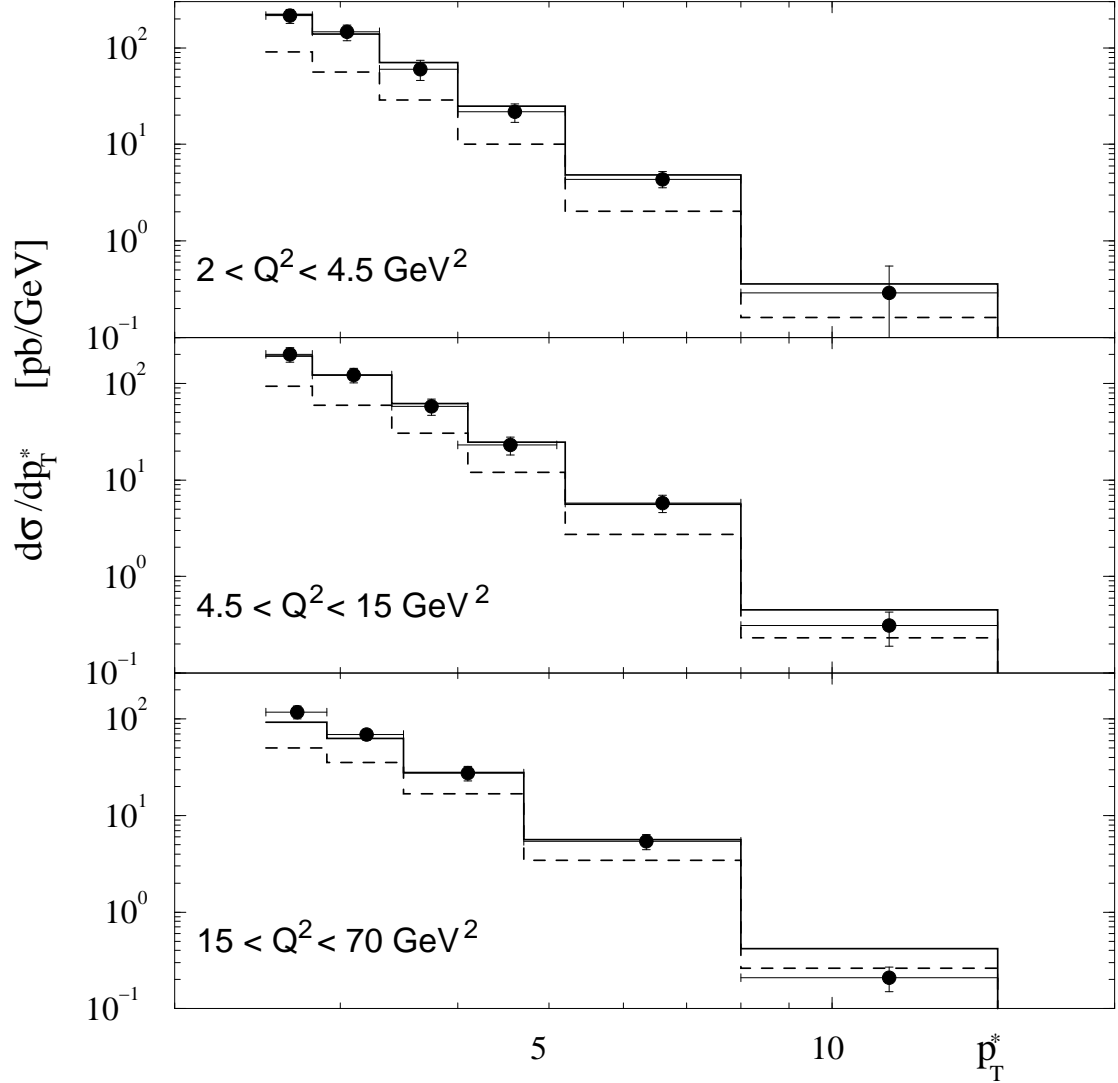}
\end{center}
\caption {Inclusive $\pi^0$ cross section as a function of the $\pi^0$
transverse momentum in the $\gamma^*$-proton center of mass frame for three
different intervals in $Q^2$. The data points are from the H1
collaboration~\cite{h1-2004}. The histograms are the theoretical results: the
solid line corresponds to full NLO predictions and the  dashed line to the 
``direct" contribution. Choice of scales is as in Fig.~\ref{fig-xbj-25}.
The variable $p^*_T$ in the figure is the notation of the H1 collaboration and
is called $E_\bot$ in the text.}
\label{fig-pt}
\end{figure}

A very impressive agreement is also achieved, in Fig.~\ref{fig-pt}, for the
$E_\bot$ spectrum for all $Q^2$ values. The resolved contribution decreases
with $Q^2$ but it remains important for all values of $Q^2$ and all transverse
momenta. The $E_\bot$ distribution should be sensitive to the choice of the
fragmentation functions and it is interesting to try different sets, in
particular that of Kretzer~\cite{kretzer}. We do not do it here as it has
already been shown by Daleo {\it et al.}~\cite{Daleo:2004pn}  that the
parametrization of ~\cite{kretzer} leads to predictions which fall below the
data. This confirms previous studies~\cite{Bourhis:2000gs} showing that the
fragmentation functions of \cite{kretzer} systematically underestimate particle
production in hadronic reactions.

\begin{figure}[!htb]
\begin{center}
\hskip -2.cm
\includegraphics[width=15cm]{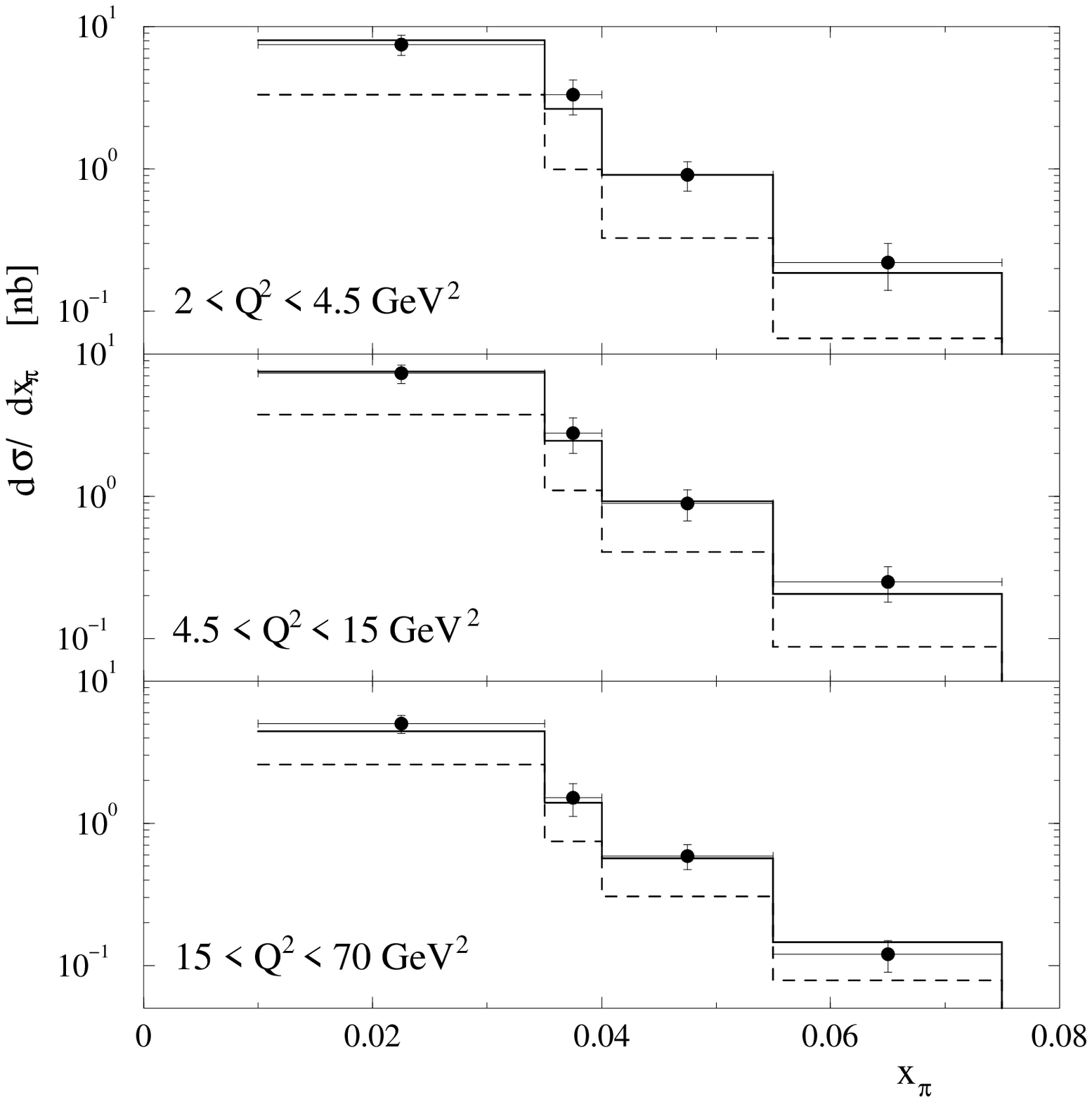}
\end{center}
\caption {Inclusive $\pi^0$ cross section as a function of $x_\pi =
E^{lab}_\pi/E^{lab}_p$ in  the range $E_\bot >$ 2.5 GeV for three different
intervals in $Q^2$. The cuts on other variables are given in the text.  The
data points are from the H1 collaboration~\cite{h1-2004}. The histograms are
the theoretical results:the solid line corresponds to full NLO predictions and
the  dashed line to the ``direct" contribution. Choice of scales is as in
Fig.~\ref{fig-xbj-25}.}
\label{fig-xpi-q2}
\end{figure}

\begin{figure}[!htb]
\begin{center}
\hskip -2.cm
\includegraphics[width=15cm]{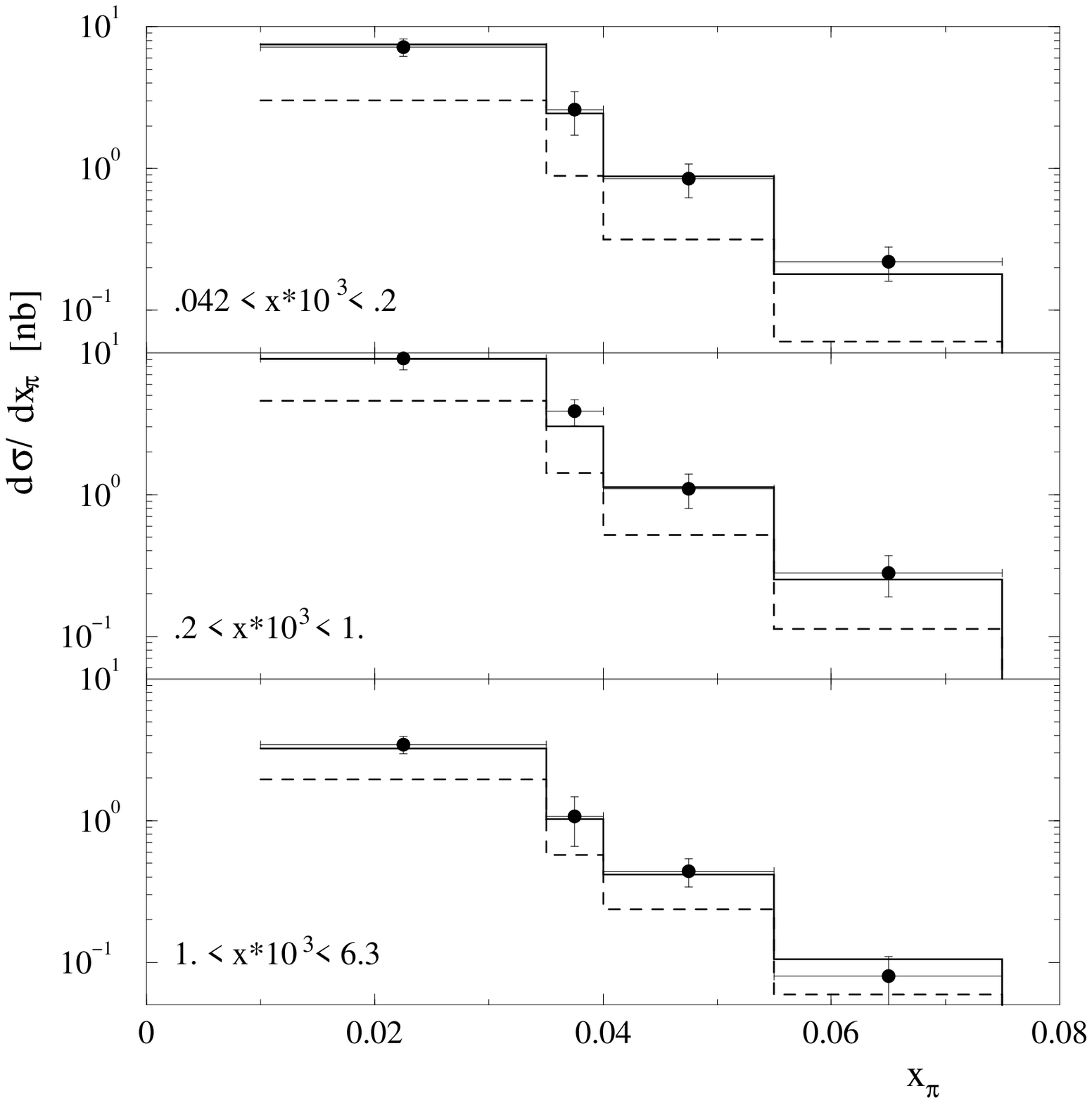}
\end{center}
\caption {Inclusive $\pi^0$ cross section as a function of $x_\pi =
E^{lab}_\pi/E^{lab}_p$ in  the range $E_\bot >$ 2.5 GeV for three different
intervals in $x_{Bj}$. The cuts on other variables are given in the text.  The
data points are from the H1 collaboration~\cite{h1-2004}. The histograms are
the theoretical results: the solid line corresponds to full NLO predictions and
the  dashed line to the ``direct" contribution. Choice of scales is as in
Fig.~\ref{fig-xbj-25}.} 
\label{fig-xpi-xbj}
\end{figure}

Similarly, the longitudinal momentum distribution of the pion is in remarkable
agreement with the data both for specific $Q^2$ bins (Fig.~\ref{fig-xpi-q2}) or specific
$x_{Bj}$ bins (Fig.~\ref{fig-xpi-xbj}). Again the resolved component
is important over the whole $x_\pi$ range but, clearly, it gives a decreasing
contribution as $Q^2$ increases. One has to note that there is little
correlation between $x_\pi$ and the fragmentation variable $z$: for instance, in
the resolved case $<\!\! z \!\!> \sim .5$ (calculated with the Born term only in
the region 2~GeV$^2 < Q^2 < $~4.5~GeV$^2$) 
varies by less than 10\% when $x_\pi$ varies between the first and the last bin
of Fig.~\ref{fig-xpi-q2}. Therefore we cannot rely on the $x_\pi$ spectrum to
constrain the $z$ shape of the fragmentation functions.

From the comparison with data we can conclude that perturbative QCD, in the NLO
approximation, gives unexpectedly good results, especially in view of the
initial discrepancy  observed at leading order between theory and data. Two
ingredients explain this fact: the unusually large correction to the direct
term, specially at low $x_{Bj}$, and the importance of the resolved photon
contribution including the associated higher order corrections. Unfortunately,
none of the inclusive observables discussed here allows for an unambiguous
separation of the two terms. In principle, looking at more exclusive
quantities, such as hadron-jet correlations, would allow the
determination of the longitudinal momentum fraction in the photon
$x_\gamma$~\cite{Fontannaz:2002nu}, and consequently the separation of the two
types of terms. However, since the HO$_s$ term is very large, it may lead to 
a large contribution at $x_\gamma \ne 1$ making the separation from the
resolved term difficult.
\begin{figure}[!htb]
\begin{center}
\hskip -2.cm
\includegraphics[width=15cm]{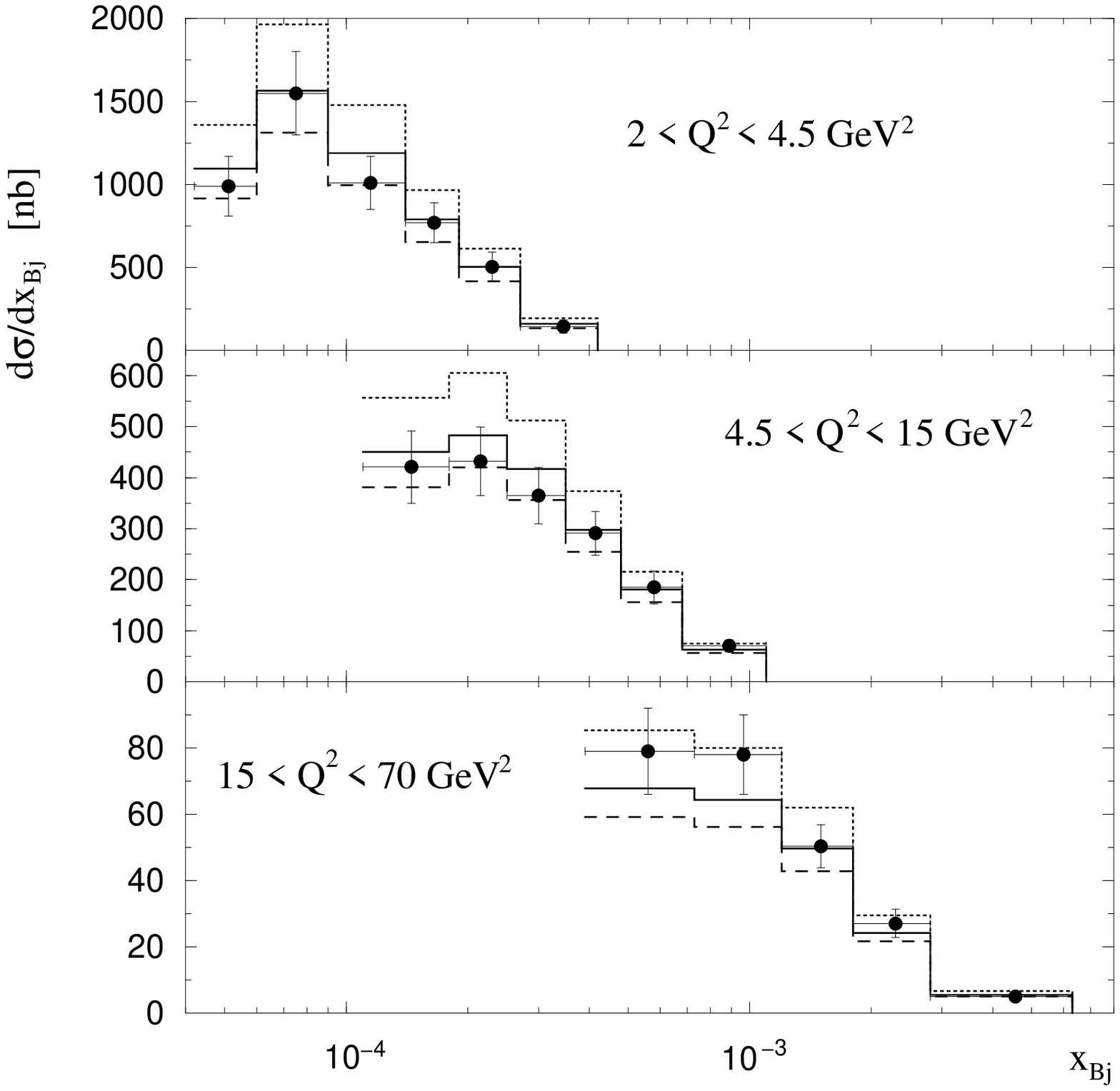}
\end{center}
\caption {Inclusive $\pi^0$ cross section as a function of $x_{Bj}$ in  the
range $E_\bot >$ 2.5 GeV for three different intervals in $Q^2$.  The data
points are from the H1 collaboration~\cite{h1-2004}. The histograms are the
NLO theoretical results for different scales of the form $C^2 (Q^2+E_\bot^2)$:
$C^2 = .5$, upper dotted histogram; $C^2 = 1$, solid histogram; $C^2 = 2$ lower dashed
histogram.} 
\label{fig-xbj-25-scale}
\end{figure}
\begin{figure}[!ht]
\begin{center}
\hskip -2.cm
\includegraphics[width=15cm]{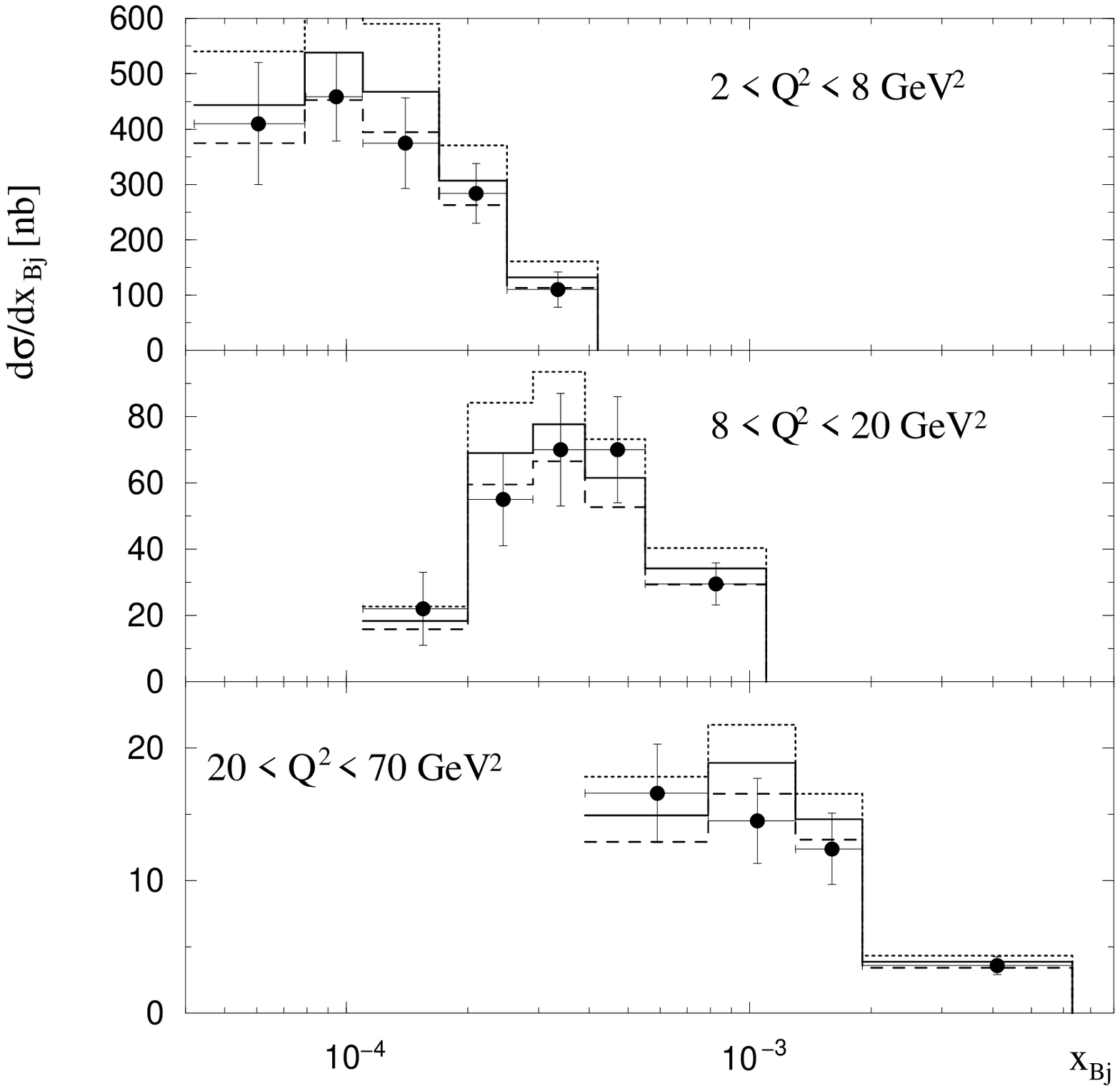}
\end{center}
\caption {Inclusive $\pi^0$ cross section as a function of $x_{Bj}$ in  the
range $E_\bot >$ 3.5 GeV for three different intervals in $Q^2$. The data points
are from the H1 collaboration~\cite{h1-2004}. The histograms are the NLO
theoretical results for different scales of the form $C^2 (Q^2+E_\bot^2)$: $C^2 =
.5$, upper dotted histogram; $C^2 = 1$, solid histogram; $C^2 = 2$ lower dashed
histogram.}
\label{fig-xbj-35-scale}
\end{figure}
 
The success of perturbative QCD to explain the data at small $x_{Bj}$ is
interesting. It seems to imply that there is no clear signal in the H1
data of the BFKL type resummation effects and 
that keeping only the lowest order term in the
usual perturbative sense is justified. One reason may be the following. The
BFKL result is derived for asymptotic energies. However, at HERA the rapidity
range, $\ln(S/Q^2)$, available is not extremely large and threshold effects do
not allow for the full formation of the BFKL
ladder~\cite{Orr:1997im,Andersen:2002zq}.

\section{Comparison with other perturbative calculations}

The H1 data are also in very good agreement with the NLO calculations of  Daleo
{\it et al.}~\cite{Daleo:2004pn} and Kniehl {\it et al.}~\cite{Kniehl:2004hf}.
We recall that the difference between these approaches and the present one 
lies in
the fact that, in the former, no special consideration is given to the
photon structure function: the NLO correction to the direct term contains a
``large" factor of type $\ln((Q^2+E_\bot^2)/Q^2)$ which amounts, in fact, to
parametrizing the photon function by its lowest perturbative approximation.
Furthermore no NLO corrections are included in the resolved cross section. In
contrast, in this work, we use both a NLO expression for the resolved photon
structure function and we include the HO corrections to the resolved cross
section. Agreement with the data is obtained in all cases at the cost of using
a different choice for the common scale. In  \cite{Daleo:2004pn} and 
\cite{Kniehl:2004hf}, as well as in our previous work~\cite{Aurenche:2003by},
the scale $(Q^2+E_\bot^2)/2$ was the appropriate choice. In this work, it is
seen that the scale $(Q^2+E_\bot^2)$ is preferred. The data do not obviously
prefer one or the other of the two sets of calculations as the shape of the
observables is not affected. In Ref. ~\cite{Kniehl:2004hf} a very
large scale sensitivity was however observed: under the rather modest change
from  $(Q^2+E_\bot^2)/4$ to $(Q^2+E_\bot^2)$ the theoretical predictions vary
by as much as a factor 2 in some cases, and, in any case, the theoretical
uncertainties are much (sometimes twice) larger that the experimental ones
(statistic and systematic errors combined).  In the  current 
approach we expect a smaller sensitivity to the scales since more HO
corrections are taken into account. Besides, it is seen from Figs.~\ref{fig:3},
\ref{fig:5}, \ref{fig:6} that the variation with the scales seems to decrease
at higher scales. This is illustrated in Figs.~\ref{fig-xbj-25-scale} and
\ref{fig-xbj-35-scale} where we show the results for $(Q^2+E_\bot^2)/2$ and $2
(Q^2+E_\bot^2)$~\footnote{Note that the resolved photon scale differs slightly
from the others since we use, as explained in Sec.~\ref{sec:theory},
$M^2_\gamma = (Q^2+ C^2_\gamma E_\bot^2)$ with $C^2_\gamma =1/2$ and 
$C^2_\gamma=2$.}. Compared to
the results of~\cite{Kniehl:2004hf}, the scale variations are somewhat
tempered and are of the same order as that of the rather large experimental 
errors.

\section{Conclusions}

Using the latest structure and fragmentation functions, the complete NLO
calculation of the direct and resolved contributions to forward particle
production in deep-inelastic scattering at HERA, describes the data rather well
in the wide kinematical range available:  2~GeV$^2 < Q^2 <$ 70~GeV$^2$,
2.5~GeV $< E_\bot <$ 15~GeV. The importance of the NLO corrections to both the
direct and resolved terms is pointed out. These large corrections are
associated with new topologies involving gluon exchange in the hard
sub-processes. These terms, which have no equivalent at the lowest order are
interpreted as the first terms of the BFKL ladder. The data seem to
indicate that resummation of such ladder diagrams is not necessary, probably
because of the not so large rapidity phase space available. Agreement between
theory and data is achieved choosing a standard scale of the form 
$(Q^2+E_\bot^2)$.  The variations under the proton factorization scale and the
photon factorization scale are under control. However a rather large
instability of the predictions is observed when varying independently the
renormalization and the fragmentation scales. This prevents a really
quantitative prediction for the single pion inclusive distribution in the
forward region. In this respect, taking account of the HO resolved contribution
improves the situation compared to calculations which ignored it but
the situation is still far from satisfactory. We have checked that imposing a 
larger
$E_\bot$ cut on the data reduces the scale sensitivity: for example, with the
conditions $E_\bot >$ 7~GeV and  4.5~GeV$^2 < Q^2 <$ 15~GeV$^2$ the variation is
$\pm^{13\%}_{10\%}$ for a scale variation as in Fig.~\ref{fig-xbj-25-scale}.
Probably, the evaluation of the next-to-next-to-leading order terms is required
to obtain reliable and stable perturbative predictions.

\section*{Acknowledgments}

The authors thank Jacek Turnau, Gudrun Heinrich, Jean-Philippe Guillet for
discussions and Roberta Shapiro for a critical reading of the manuscript.
R.M.G. would like to thank the Department of Science and Technology, India, for
financial support to the Centre for High Energy Physics, IISc, for a cluster,
under the FIST program : SR/FIST/PSI-022/2000. P.A. and M.F. thank the Institute
of Mathematical Sciences, Chennai for hospitality and  R.M.G thanks LAPTH, 
Annecy for hospitality in addition.

\end{document}